\documentclass[a4paper,11pt]{article}

\usepackage[a4paper,
            bindingoffset=0in,
            left=2.8 cm,,
            right=2.8 cm,
            top=1.25in,
            bottom=1.25in,
            footskip=.25in]{geometry}

\usepackage[ruled,vlined,linesnumbered]{algorithm2e}

\usepackage{authblk}
\usepackage{caption}
\usepackage{subcaption}

\usepackage[utf8]{inputenc}
\usepackage[commandnameprefix=ifneeded,commentmarkup=uwave]{changes}
\usepackage[n,landau,notions,ff,mm]{cryptocode}
\usepackage{xcolor}
\usepackage{graphicx}
\definecolor{gamechangecolor}{gray}{0.74}
\usepackage{tikz}
\usetikzlibrary{arrows,positioning,decorations.pathreplacing}
\usepackage{url}
\usetikzlibrary{shapes}
\usepackage{import}
\usepackage{xifthen}
\usepackage{pdfpages}
\usepackage{transparent}
\usepackage{balance}
\usepackage{xspace}
\usepackage{amsmath}
\usepackage{hyperref}
\usepackage{cleveref}
\usepackage{amsfonts}%
\usepackage{multirow}
\usepackage{graphicx}
\usepackage{booktabs}
\usepackage{enumitem}
\setlist[itemize]{left=0pt}
\setlist[enumerate]{left=0pt}
\usepackage{makecell}
\usepackage{colortbl}
\definecolor{gainsboro}{rgb}{0.86, 0.86, 0.86}
\newcommand{\greycell}[1]{\cellcolor{gainsboro}{#1}}

\usepackage{tikz-qtree}
\usepackage{tikz}
\usepackage{pgf-umlsd}
\usepackage{dashbox}

\usepgflibrary{arrows} 
\usepackage{pgfplots}


\newcommand{\etal}{\textit{et al.}\xspace}

\definecolor{forest}{rgb}{0.0, 0.5, 0.0}

\definecolor{Gray}{gray}{0.9}

\usepackage{amssymb}
\usepackage{pifont}

\newtheorem{findings}{Finding}
\newcommand{\insightbox}[1]{
\vspace{0.5em}
\noindent \fcolorbox{black}{gray!20}{
\begin{minipage}{0.95\columnwidth}
\begin{findings}
#1
\end{findings}
\end{minipage}
}
\vspace{0.5em}
}

\usepackage{acro}
\acsetup{single}

\DeclareAcronym{DeFi}{
  short = DeFi,
  long  = Decentralized Finance,
}
\newcommand{\DeFi}{\ac{DeFi}\xspace}

\DeclareAcronym{PoW}{
  short = PoW,
  long  = Proof-of-Work,
}
\newcommand{\PoW}{\ac{PoW}\xspace}

\DeclareAcronym{PoS}{
  short = PoS,
  long  = Proof-of-Stake,
}
\newcommand{\PoS}{\ac{PoS}\xspace}

\DeclareAcronym{LTVV}{
  short = LTV,
  long  = Loan-to-Value,
}
\newcommand{\LTVV}{\ac{LTVV}\xspace}

\DeclareAcronym{HF}{
  short = HF,
  long  = Health Factor,
}
\newcommand{\HF}{\ac{HF}\xspace}
\newcommand{\HFs}{\acp{HF}\xspace}

\DeclareAcronym{LT}{
  short = LT,
  long  = Liquidation Threshold,
}
\newcommand{\LT}{\ac{LT}\xspace}

\DeclareAcronym{APR}{
  short = APR,
  long  = Annual Percentage Rate,
}
\newcommand{\APR}{\ac{APR}\xspace}

\DeclareAcronym{ROI}{
  short = ROI,
  long  = Return On Investment,
}

\DeclareAcronym{NO}{
  short = NO,
  long  = Node Operator,
}

\newcommand{\NOs}{\acp{NO}\xspace}

\DeclareAcronym{LSD}{
  short = LSD,
  long  = Liquid Staking Derivative,
}
\newcommand{\LSD}{\ac{LSD}\xspace}
\newcommand{\LSDs}{\acp{LSD}\xspace}

\DeclareAcronym{LST}{
  short = LST,
  long  = Liquid Staking Tokens,
}

\DeclareAcronym{DEX}{
  short = DEX,
  long  = Decentralized Exchange,
}
\newcommand{\DEX}{\ac{DEX}\xspace}
\newcommand{\DEXes}{\acp{DEX}\xspace}

\DeclareAcronym{CEX}{
  short = CEX,
  long  = Centralized Exchange,
}
\newcommand{\CEX}{\ac{CEX}\xspace}
\newcommand{\CEXes}{\acp{CEX}\xspace}

\DeclareAcronym{NFT}{
  short = NFT,
  long  = Non-fungible Token,
}

\DeclareAcronym{LP}{
  short = LP,
  long  = Liquidity Provider,
}

\DeclareAcronym{P2P}{
  short = P2P,
  long  = peer-to-peer,
}
\newcommand{\PtP}{\ac{P2P}\xspace}

\DeclareAcronym{TVL}{
  short = TVL,
  long  = Total Value Locked,
}
\newcommand{\TVL}{\ac{TVL}\xspace}

\DeclareAcronym{DApp}{
  short = DApp,
  long  = Decentralized Application,
}

\DeclareAcronym{DAO}{
  short = DAO,
  long  = Decentralized Autonomous Organization,
}
\newcommand{\DAO}{\ac{DAO}\xspace}

\DeclareAcronym{CeFi}{
  short = CeFi,
  long  = Centralized Finance,
}

\DeclareAcronym{MEV}{
  short = MEV,
  long  = Maximal Extractable Value,
}
\newcommand{\MEV}{\ac{MEV}\xspace}

\DeclareAcronym{AMM}{
  short = AMM,
  long  = Automated Market Maker,
}

\DeclareAcronym{FaaS}{
  short = FaaS,
  long  = Front-running as a Service,
}

\DeclareAcronym{SaaS}{
  short = SaaS,
  long  = Staking as a Service,
}
\newcommand{\SaaS}{\ac{SaaS}\xspace}

\newcommand{\USD}{\ensuremath{\xspace\texttt{USD}}\xspace}

\newcommand{\stETH}{\ensuremath{\xspace\texttt{stETH}}\xspace}

\newcommand{\ETH}{\texttt{ETH}\xspace}

\newcommand{\LidoStartCrawlingDate}{Dec~$17$,~2020\xspace}
\newcommand{\LidoEndCrawlingDate}{Aug~$7$,~2023\xspace}
\newcommand{\TerraCrashStartDate}{May~$8$,~2022\xspace}
\newcommand{\TerraCrashStartDateAndTen}{May~$18$,~2022\xspace}

\newcommand{\totalLeverageStakingAmt}{\ensuremath{\xspace 295{,}243}\xspace}
\newcommand{\totalLeverageStakingUSD}{\ensuremath{\xspace 482}m\xspace}

\newcommand{\principal}{$S$}
\newcommand{\LTV}{$l$}

\newcommand{\LEM}{$LevM_{(S,n)}$}

\newcommand{\Asset}{$A_{(S,n)}$}
\newcommand{\Collateral}{$C_{(S,n)}$}
\newcommand{\Borrow}{$B_{(S,n)}$}

\newcommand{\UST}{$\texttt{UST}$\xspace}
\newcommand{\LUNA}{$\texttt{LUNA}$\xspace}
\newcommand{\bETH}{$\texttt{bETH}$\xspace}

\newcommand{\user}{$\mathsf{U_i}$\xspace}
\newcommand{\GroupLV}{$G_{L}$\xspace}

\newcommand{\GroupOD}{$G_{O}$\xspace}

\newcommand{\oxd}{\href{https://etherscan.io/address/0xd275e5cb559d6dc236a5f8002a5f0b4c8e610701}{0xD2...701}\xspace}

\newcommand{\oxA}{\href{https://etherscan.io/address/0xA1175a219dac539F2291377F77afD786D20e5882}{0xA1...882}\xspace}

\title{Leverage Staking with Liquid Staking Derivatives (LSDs): Opportunities and Risks}

\author[1]{Xihan Xiong}
\author[1]{Zhipeng Wang}
\author[2]{Xi Chen}
\author[1]{William Knottenbelt}
\author[1]{Michael Huth}

\affil[1]{Imperial College London, UK}
\affil[2]{University of Sussex, UK}
\date{} 

\begin{document}
\maketitle

\begin{abstract}
In the Proof of Stake (PoS) Ethereum ecosystem, users can stake \ETH on Lido to receive \stETH, a Liquid Staking Derivative (LSD) that represents staked \ETH and accrues staking rewards. LSDs improve the liquidity of staked assets by facilitating their use in secondary markets, such as for collateralized borrowing on Aave or asset exchanges on Curve. The composability of Lido, Aave, and Curve enables an emerging strategy known as leverage staking, an iterative process that enhances financial returns while introducing potential risks. This paper establishes a formal framework for leverage staking with \stETH and identifies $442$ such positions on Ethereum over $963$ days. These positions represent a total volume of $537{,}123$ \ETH ($877$m USD). Our data reveal that $81.7\%$ of leverage staking positions achieved an Annual Percentage Rate (APR) higher than the APR of conventional staking on Lido. 
Despite the high returns, we also recognize the potential risks. For example, the Terra crash incident demonstrated that token devaluation can impact the market. Therefore, we conduct stress tests under extreme conditions of significant \stETH devaluation to evaluate the associated risks. 
Our simulations reveal that leverage staking amplifies the risk of cascading liquidations by triggering intensified selling pressure through liquidation and deleveraging processes. Furthermore, this dynamic not only accelerates the decline of stETH prices but also propagates a contagion effect, endangering the stability of both leveraged and ordinary positions.


\end{abstract}

\section{Introduction}

The Ethereum blockchain's transition from \PoW~\cite{wood2014ethereum} to \PoS\cite{grandjean2023ethereum,schwarz2022three,agrawal2022proofs,tang2023transaction} is a remarkable shift towards a more sustainable consensus mechanism. This change, while crucial for energy efficiency, introduces new challenges in staking \ETH. In \PoS-based Ethereum, validators must stake \ETH to secure the network~\cite{buterin2020combining,neu2021ebb} and earn staking rewards. However, solo staking demands a substantial capital commitment of $32$ \ETH and technical expertise in maintaining a validator node. Additionally, staked \ETH becomes illiquid during the staking period, limiting its usability for other financial activities.

To mitigate these challenges, \LSDs, also referred to as Liquid Staking Tokens (LSTs),  have emerged as transformative solutions. These assets enhance the liquidity of staked assets while preserving their earning potential. Retail users can flexibly stake any amount of \ETH on a liquid staking platform (e.g., Lido) to receive the corresponding \LSDs. These \LSDs are fungible and tradable representations of the staked \ETH and its associated rewards. At the time of writing, \href{https://lido.fi/}{Lido} stands as a leading \LSD provider on Ethereum, marked by its top position with a \TVL of $38$b~\USD
\footnote{\url{https://defillama.com/protocol/lido}, last accessed on Dec~$5$,~$2024$.}.

In the LSD primary market, platforms such as Lido allow users to convert \ETH into \stETH for various uses within the \DeFi ecosystem. Specifically, users might choose to simply hold \stETH to accrue a staking \APR of around $3.6\%$ (as of Mar 2024) or utilize \stETH in the secondary market for further financial activities. For example, \stETH can serve as collateral on \DeFi lending platforms such as \href{https://aave.com/}{Aave} to borrow \ETH. This allows users to earn rewards on their staked \ETH while utilizing \stETH as active investment capital\footnote{\url{https://github.com/lidofinance/aave-asteth-deployment}}. Additionally, \stETH can be used for liquidity provision and traded for \ETH in the \stETH--\ETH pool of a \DEX, such as the \href{https://curve.fi/#/ethereum/swap}{Curve} protocol.

The composability between Lido, Aave, and Curve facilitates two novel strategies of leverage staking (see Figure~\ref{fig: lev_pic_direct} and~\ref{fig: lev_pic_indirect}). The first, known as \emph{direct leverage staking}, involves users staking \ETH on Lido in the primary market to receive \stETH, which is then used as collateral on Aave to borrow \ETH, subsequently restaked on Lido. Users can iteratively execute this process to increase financial returns based on their risk profile. The second strategy, \emph{indirect leverage staking},  involves initially swapping \ETH for \stETH within the Curve pool at secondary market prices, then using the acquired \stETH as collateral on Aave to borrow more \ETH, which is swapped again for \stETH in the Curve pool. This allows users to participate in staking and earn rewards without directly staking their \ETH on Lido. Together, these strategies demonstrate the flexibility and depth of the \LSD ecosystem, offering varied approaches to increasing returns with leveraged positions.

While leverage staking offers high return opportunities, it also presents potential risks. Under adverse market conditions that lead to a substantial decline in \stETH prices, leverage staking can act as a catalyst for market instability by increasing the risk of \emph{cascading liquidations}, where successive liquidations drive \stETH prices into a downward spiral. As such, this paper aims to understand the opportunities and risks of leverage staking. We investigate the underlying mechanisms, evaluate the financial benefits and inherent risks of this strategy, and assess its broader market impact. We outline our main contributions as follows:
\begin{description} [leftmargin=*,itemsep=0.05em]
    \item[Strategy Formalization.] We develop a formal framework for leverage staking with \stETH that captures direct and indirect strategies. We conduct an analytical study to derive key metrics such as leverage staking multiplier, \HF, and APR for each position. To the best of our knowledge, we are the first to model leverage staking with \LSDs.

    \item[Empirical Measurement.] We empirically analyze leverage staking spanning $963$ days, from Dec~$17$,~$2020$ to Aug~$7$,~$2023$. We detect $262$ direct leverage staking positions with a total stake amount of \totalLeverageStakingAmt~\ETH (\totalLeverageStakingUSD~\USD), and $180$ indirect leverage staking positions with a total swap amount of $241{,}880$~\ETH ($395$m~\USD). We observe that a majority ($81.7\%$) of leverage staking positions yielded an \APR higher than conventional staking on Lido.

    \item[User Behavior Analysis.]  We explore the \stETH price deviation in the context of the Terra crash incident. We analyze how users behave when faced with potential liquidations. We discover that users actively deleveraged their leverage staking positions and collectively repaid a substantial debt of $136{,}069$~\ETH, further intensifying the \stETH selling pressure. 

    \item[Stress Testing.] We perform stress tests on the Lido--Aave--Curve \LSD ecosystem to evaluate the impact of leverage staking under extreme conditions of significant \stETH devaluation. Our simulation reveals that leverage staking escalates the risk of cascading liquidations. Systemic risk is exacerbated not only through liquidations but also via the market pressures these actions generate. The selling pressure on \stETH, driven by both liquidations and deleveraging actions, can trigger a contagion effect across the system, further depressing \stETH prices and adversely impacting the stability of broader market participants, including ordinary users. 
\end{description}

\section{Related Work}
We provide an overview of the literature on \PoS staking, \LSDs, and \DeFi lending.

\smallskip\emph{\PoS Staking.} The economics of \PoS staking has been explored in prior research works. For example, Cong~\etal~\cite{cong2022staking} developed a continuous-time model to explore the economic impact of staking in token-based digital economies. They found that higher staking rewards lead to increased staking ratios, which in turn predict higher token price appreciation and generate profitable carry trade opportunities with significant Sharpe ratios. Additionally, attention has been given to the security of PoS staking.  For instance, Chitra~\cite{chitra2021competitive} investigated how on-chain lending affects the security of a \PoS blockchain. They found that when the yield from lending contracts is higher than the inflation rate from staking, stakers are incentivized to remove their staked tokens and lend them out, thus reducing network security.

\emph{\LSDs.} Tzinas~\etal~\cite{tzinas2023principal} studied the Principal-Agent problem in the liquid staking setting. They discussed the dilemma between the choice of proportional representation and fair punishment and proposed a concrete attack to illustrate their incompatibility. Scharnowski~\etal~\cite{scharnowski2023economics} analyzed the liquid staking basis (e.g., the discrepancy) between the prices of \LSDs in the primary and secondary market. They observed that the liquid staking basis widens when cryptocurrency volatility increases and liquidity decreases in the secondary market. Cintra~\etal~\cite{cintra2023detecting} utilized the Bayesian Online Changepoint Detection algorithm to identify potential depeg incidents using price data from the Curve \stETH--\ETH pool. They show that the proposed approach can help users manage potential risks.

\smallskip\emph{\DeFi Lending.} Heimbach~\etal~\cite{heimbach2023defi} studied the impact of the Ethereum merge on two \DeFi lending platforms, Compound and Aave. They investigated the actions taken by Aave to mitigate the liquidation risk of collateralized \stETH positions. Wang~\etal~\cite{wang2022speculative} formalized a model for under-collateralized \DeFi lending platforms and empirically evaluated the risks associated with leveraging, such as impermanent loss, arbitrage and liquidation.

\section{Background}

\subsection{Blockchain and DeFi}

Blockchain is a decentralized digital ledger that records transactions across multiple nodes to ensure security, transparency, and immutability. It is the foundational technology behind cryptocurrencies such as Bitcoin~\cite{bitcoin} and Ethereum~\cite{wood2014ethereum}, enabling \PtP transactions without the need for a trusted third party. The blockchain structure as a series of blocks chained together through cryptographic hashes helps prevent alteration to the data once it has been confirmed on-chain. A permissionless blockchain allows any participant to join and engage without requiring authorization. In this context, the Ethereum blockchain~\cite{buterin2013ethereum} emerges as a pioneering platform, supporting the execution of smart contracts and empowering developers to create various decentralized applications.

\DeFi~\cite{werner2022sok}  represents an innovative application of blockchain technology, focusing on building open financial systems. \DeFi refers to a set of blockchain-based financial services and products that operate without intermediaries, using smart contracts to build an open environment. \DeFi innovations ranging from collateralized lending to \DEXes are reshaping the financial system. The \TVL in \DeFi hit a record high of $178$b \USD in Nov~$2021$, with the Ethereum blockchain driving these \DeFi activities.

\subsection{Ethereum PoS}\label{sec: intro_pos}
    
The \PoS consensus mechanism, first proposed in online forums and later examined by academia~\cite{daian2019snow,buterin2020combining,gavzi2019proof,kiayias2017ouroboros,bano2019sok}, has emerged as an energy-efficient alternative to \PoW. 

\smallskip\emph{Beacon Chain.} On Dec~$1$,~$2020$, Ethereum marked a significant milestone by introducing its PoS-based Beacon Chain that runs in parallel with Ethereum's \PoW Mainnet. In the Beacon Chain,  ``staking'' is introduced through a deposit mechanism, allowing participants to become validators by locking up $32$ \ETH in a designated smart contract. Staking enables validators to contribute to the integrity of the network by participating in the consensus process, including proposing and validating blocks. This not only helps maintain the security of the blockchain, but also allows stakeholders to earn rewards proportional to their contributions, incentivizing more participants to engage in the network's operation.

\smallskip\emph{The Merge.}  On Sep~$15$,~$2022$, the Merge enables Beacon Chain to evolve as the consensus mechanism for the entire Ethereum network~\cite{EthMerge2022}. Ethereum now runs on the execution layer and the consensus layer. The execution layer is responsible for executing transactions, defining how the state of the Ethereum network changes over time. The role of the consensus layer entails establishing agreement among validators regarding the state of the execution layer. The Ethereum staking system offers various incentives to validators. Rewards from the consensus layer include block proposal, attestation, and sync committee rewards~\cite{grandjean2023ethereum}. The execution layer introduces additional rewards, including priority tips and \MEV tips~\cite{daian2020flash,liu2022empirical}. Penalties also apply to dishonest behaviors.

\smallskip\emph{The Shapella Upgrade.} On Apr~$12$,~$2023$, Ethereum underwent the ``Shapella upgrade''. The Shapella upgrade combines the ``Shanghai upgrade'' and the ``Capella upgrade'', which took place on the consensus and execution layer simultaneously~\cite{EthHistory2023}. The Shapella upgrade primarily introduces the capability to unstake \ETH secured within the network.  This ability enhances the operational dynamics for both individual stakers and validators. Validators can initiate withdrawals of their staked \ETH, either partially or in full, enabling them to reclaim their capital and potentially redeploy it elsewhere. Similarly, if a user has staked \ETH on Lido~(see Section~\ref{sec:lsd_intro}), they now have the flexibility to partially or fully unstake their assets, allowing for greater liquidity and control over their investments.

\subsection{Staking Options}
Ethereum participants are presented with four distinct staking options as follows.
\begin{description}[leftmargin=*, itemsep=0.1em]
\item[Solo Staking.] In solo staking, individual participants operate their validator nodes by committing a threshold of $32$ \ETH, thus maintaining full control over the staking rewards. However, solo staking necessitates technical expertise to manage a validator node. Furthermore, its substantial capital requirement may render it financially inaccessible for many retail users.

\item[\SaaS.] For users possessing the requisite $32$ \ETH but lacking in technical expertise, \SaaS presents a viable solution. \SaaS manages the validator node on behalf of users, utilizing their signing keys to perform on-chain tasks\footnote{\url{https://ethereum.org/en/staking/saas/}}. This simplifies the staking process for individuals and mitigates the risks associated with node management.

\item[Pooled Staking.] For retail users with holdings below the $32$ \ETH threshold, pooled staking emerges as a feasible alternative, enabling them to collectively participate in the network's validation process, earn rewards, and capitalize on the broader Ethereum ecosystem without the need for individual, full-node commitments. Typically, staking pools charge fees, which are further split between \NOs and the protocol \DAO. \NOs run and maintain validator nodes on behalf of the staking pool, while the \DAO selects \NOs and configures crucial parameters for the protocol.

\item[\CEX Staking.] \CEXes, such as \href{https://www.coinbase.com/earn}{Coinbase} and \href{https://www.binance.com/en/defi-staking}{Binance}, provide centralized and custodial staking services to users.  These services simplify the staking process by managing technical aspects and providing a user-friendly interface. However, such convenience comes with inherent risks associated with the centralized nature of \CEX staking. Users must trust these platforms with their assets, making them vulnerable to security breaches, regulatory changes, or operational failures.
\end{description}

\subsection{LSD}\label{sec:lsd_intro}

\begin{figure}[tbh]
    \centering
    \includegraphics[width=0.95\columnwidth]{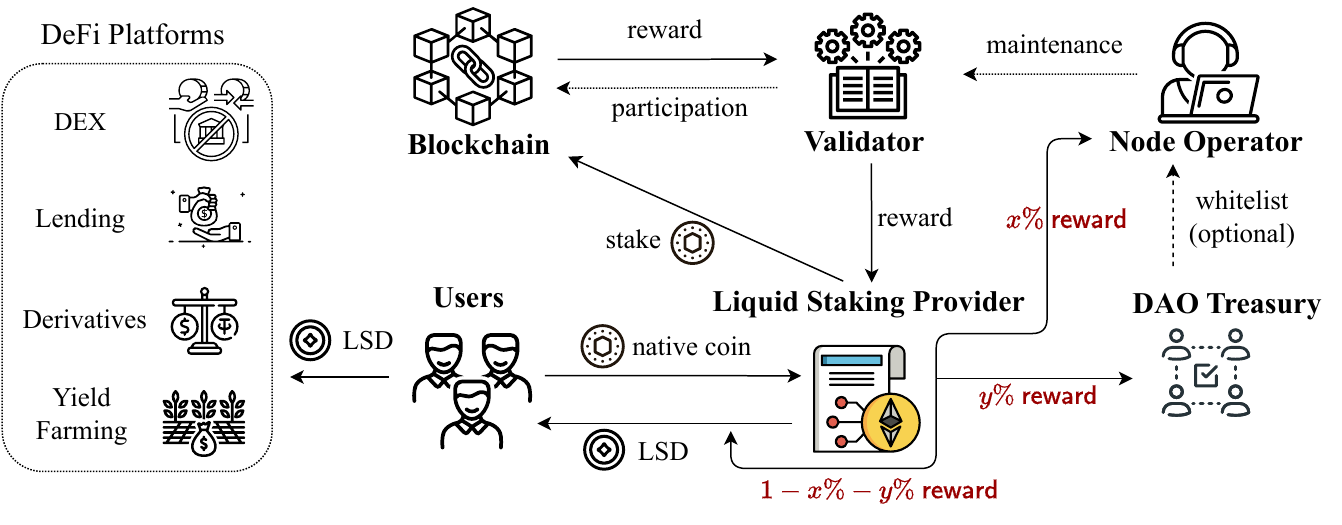}
    \caption{Overview of the LSD Ecosystem.}
    \label{fig: lsd}
\end{figure}

Staking offers several advantages, from earning rewards to enhancing network security. However, once \ETH is locked for staking, it becomes illiquid, making it inaccessible for trading. Given this challenge, the concept of \LSD emerged, which represents staked assets and rewards in a tradable form. Figure~\ref{fig: lsd} provides an overview of the \LSD ecosystem. When users stake \ETH within an \LSD provider (e.g., a liquid staking pool), they receive \LSDs in return. 

At the time of writing, liquid staking protocols accumulate a \TVL of more than $65$b \USD\footnote{\url{https://defillama.com/protocol/lido}, last accessed on Dec 5, 2024}, securing the top position in \TVL across various \DeFi sectors. Users can obtain \LSDs through two primary staking methods: pooled staking and \CEX staking. Pooled staking protocols such as \href{https://lido.fi/}{Lido}, \href{https://rocketpool.net/}{Rocket Pool}, \href{https://frax.finance/}{Frax}, \href{https://stakewise.io/}{Stakewise}, and \href{https://www.swellnetwork.io/}{Swell Network} provide \LSDs to users. \CEXes such as \href{https://www.coinbase.com/}{Coinbase} and \href{https://www.binance.com/en}{Binance} also support \LSDs. 
    
Lido is currently the leading \LSD provider and ranks as the largest \DeFi protocol in terms of \TVL. \href{https://lido.fi/}{Lido} is recognized as the leading \LSD provider and ranks as the largest \DeFi protocol in terms of \TVL. The total amount of \ETH staked on Lido reached $9.67$m in Dec~$2024$, accounting for $70.5\%$ of the total \ETH \LSDs on Ethereum.  Users stake \ETH on Lido to receive \stETH in return. \stETH implements the \emph{rebasing mechanism}, where \stETH holders' account balances get adjusted daily to reflect the accumulated rewards~\cite{lido-tokens-integration-guide2023}. The rebase can be positive or negative, depending on the validators' performance. 





\subsection{DeFi Lending Protocols} \label{aave_intro}

 \DeFi lending protocols are decentralized platforms that facilitate \PtP lending and borrowing of cryptocurrency assets through the automated execution of smart contracts. At the time of writing, Aave stands as the leading \DeFi lending protocol, with a total \TVL of $36$b \USD\footnote{\url{https://defillama.com/protocol/aave}, last accessed on Dec~$5$,~$2024$.}. The Aave V2 lending protocol follows an over-collateralization model, meaning that users must supply more collateral value than the borrowed amount. As an illustration, when the collateral value amounts to $S$ \ETH, the user's borrowing capacity is restricted to no more than $S \cdot l$ \ETH, where $l\in[0,1]$ denotes the \LTVV ratio. In the event that the collateral value falls below a specified threshold, users may need to add more collateral or risk the liquidation of their asset to repay the borrowed amount and accrued interest. To monitor the collateralization status of each position, Aave utilizes \LT to establish the threshold percentage that designates a position as under-collateralized, and \HF as a key metric to quantify the liquidation status of a position. Specifically, user \user's position can be liquidated if $\mathit{HF}_{(\mathsf{U_i})}<1$ (see Equation~\ref{HF_general}).
 \begin{equation}\label{HF_general}
   {\small
     \begin{split}
         \mathit{HF}_{(\mathsf{U_i})} = \frac{\sum_j \text{collateralized }\text{value of } \text{asset}_{j}\text{ in }\texttt{ETH}\cdot \text{LT}_{j} }{\sum_j\text{borrowed }\text{value of } \text{asset}_{j}\text{ in }\texttt{ETH}} 
     \end{split}
     }
 \end{equation}

\section{System Model}

We first provide a set of notations to facilitate the understanding of related equations. 

\begin{table}[tbh]
    \centering
    \renewcommand\arraystretch{1.5}
    \resizebox{\columnwidth}{!}{
    \begin{tabular}{ll|ll}
    \toprule
     \greycell{Notation} & Meaning& \greycell{Notation}& Meaning\\
     \midrule
    \greycell{S}	&	Initial investment (principal amount) in \ETH	&	\greycell{$p_{t_0}^{1st}$}	&	\stETH price in the primary market at $t_0$	\\
    \greycell{$l$}	&	The Loan-to-Value (LTV) ratio used by Aave	&	\greycell{$p_{t_0}^{2nd}$}	&	\stETH price in the secondary market at $t_0$	\\
    \greycell{$\text{LT}$}	&	The liquidation threshold used by Aave	&	\greycell{$p_{t_0}^{a}$}	&	\stETH price used by Aave at time $t_0$	\\
    \greycell{$\mathit{HF}_{\mathsf{U_i}}$}	&	The health factor (HF) of $U_i$'s position	&	\greycell{$p_{t_c}^{a}$}	&	\stETH price used by Aave at time $t_c$	\\
    \bottomrule
    \end{tabular}
    }
    \caption{Notations used to formalize the leverage staking strategy.}
    \label{tab:my_label}
\end{table}

\subsection{System Participants}

We consider an \LSD ecosystem on the Ethereum blockchain with the following participants. 
\begin{description} [leftmargin=*]
    \item[Users:] A user (\user) is depicted as a rational and strategic entity, proficient in interacting with multiple \DeFi platforms. \user can adopt diverse strategies to maximize financial returns.

    \item[Liquid Staking Providers:] \user can stake \ETH on liquid staking platforms (e.g., Lido) to receive \LSDs (e.g., \stETH). These \LSDs can then be used in various financial activities, including trading, collateralized borrowing, and liquidity provision.

    \item[Lending and Borrowing Providers:] \user can supply a single asset on \DeFi lending platforms such as Aave, using it as collateral to secure a loan in the form of a different asset.
\end{description}

\subsection{Leverage Staking with LSDs}
This section introduces and compares three strategies: \emph{(i)} leverage borrowing, \emph{(ii)} direct leverage staking, and \emph{(iii)} indirect leverage staking strategies.

\begin{table}[tbh]
    \centering
    \renewcommand\arraystretch{1.5}
    \resizebox{\columnwidth}{!}{
    \begin{tabular}{l|ccccc}
         \toprule
         \makecell{Investment\\Strategy} & \makecell{Asset\\Pair} & \makecell{Leverage\\Pattern} & \makecell{Staking\\Reward} & \makecell{Deposit-Borrow\\Revenue}\\
         \midrule
         \greycell{Leverage Staking (Direct)} & \ETH--LSD & Stake$\rightarrow$Deposit$\rightarrow$Borrow$\rightarrow$Stake & \ding{52}& \ding{52}   \\
        \greycell{Leverage Staking (Indirect)} & \ETH--LSD & Swap$\rightarrow$Deposit$\rightarrow$Borrow$\rightarrow$Swap & \ding{52}& \ding{52}   \\
        \greycell{Leverage Borrowing} & Non-LSD Pair & Swap$\rightarrow$Deposit$\rightarrow$Borrow$\rightarrow$Swap & \ding{54}& \ding{52}   \\
         
         \bottomrule
    \end{tabular}
    }
    \caption{Comparison of leverage staking and leverage borrowing strategies on Ethereum.}
    \label{tab:lev_stake_vs_borrow}
\end{table}

\subsubsection{Leverage Borrowing}

In Etherem, \emph{leverage borrowing} emerges as a commonly adopted strategy. This process involves users initially exchanging asset X for Y within a \DEX pool. Subsequently, asset Y is supplied as collateral on lending platforms to borrow asset X. This borrowed asset X is then exchanged again for Y in the \DEX pool, enabling users to iteratively amplify their leverage borrowing positions. The financial incentive behind the leverage borrowing strategy lies in the user's ability to expand the deposit-borrow leverage through repeated swap, deposit, and borrow actions. Although the deposit rate offered by the lending platform is lower than the borrowing rate, the larger total amount of asset Y deposited compared to asset X borrowed typically results in net earnings, which can be further amplified through leverage.

\subsubsection{Leverage Staking}

\begin{figure}[tbh]
\centering
\begin{minipage}{.95\textwidth}
  \centering
  \includegraphics[width =\columnwidth]
  {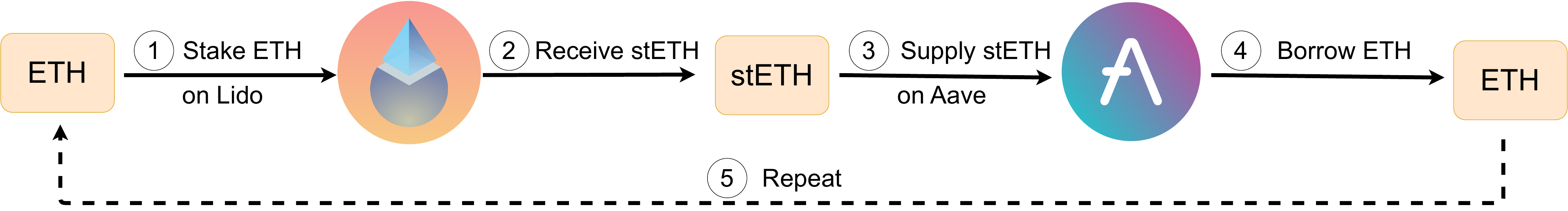}
  \caption{Overview of the direct leverage staking strategy.}
  \label{fig: lev_pic_direct}
\end{minipage}%

\smallskip
\begin{minipage}{.95\textwidth}
  \centering
  \includegraphics[width =\columnwidth]
  {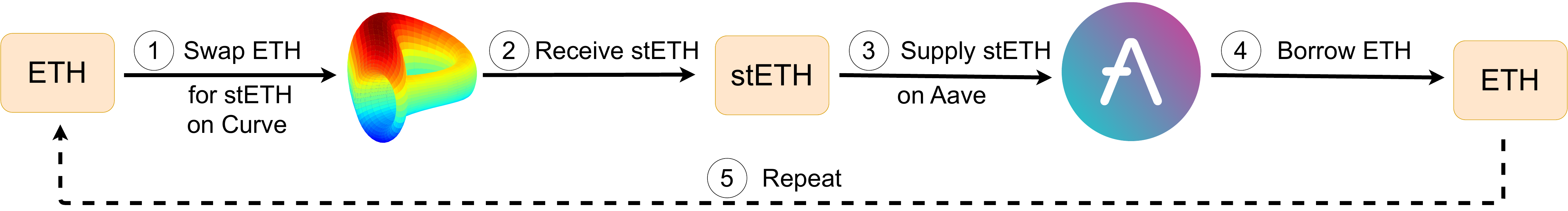}
  \caption{Overview of the indirect leverage staking strategy.}
  \label{fig: lev_pic_indirect}
\end{minipage}
\end{figure}

Following the introduction of \PoS staking, a concept analogous to leverage borrowing, termed \emph{leverage staking}, has emerged. It is a strategy linked with \LSDs and involves a recursive cycle of stake/swap, deposit, and borrow actions (see Figure~\ref{fig: lev_pic_direct} and~\ref{fig: lev_pic_indirect}) to increase financial returns. We concentrate our analysis on leverage staking within the Lido--Aave--Curve ecosystem and describe direct and indirect leverage staking as follows.

\begin{description} [leftmargin=*, itemsep=0.1em]
\item[Direct Leverage Staking.] \user first stakes a principal amount of \principal~\ETH on Lido at time $t_0$ to acquire $S/p_{t_0}^{1st}=S$~\stETH, where $p_{t_0}^{\text{1st}}=1$ denotes the \stETH to \ETH price in the primary market. Next, \user supplies \stETH on Aave to borrow $S\cdot l \cdot p_{t_0}^{a}$~amount of \ETH, where \LTV~denotes the \LTVV ratio and $p_{t_0}^{a}$~denotes the \stETH to \ETH price used by Aave V2 lending protocol\footnote{The Aave V2 lending protocol uses the Chainlink price oracle. See~\url{https://etherscan.io/address/0xa50ba011c48153de246e5192c8f9258a2ba79ca9\#code}.}. Then \user restakes the borrowed \ETH on Lido. \user performs this \emph{loop} for $n$ times, amplifying both the staking reward from Lido and the deposit-borrow revenue from Aave.

\item[Indirect Leverage Staking.]  Instead of acquiring \stETH from Lido, \user can first swap \principal~\ETH for $S/p_{t_0}^{\text{2nd}}$~\stETH\footnote{Ignore swap fees for illustration purposes.} within the Curve pool at time $t_0$, where $p_{t_0}^{\text{2nd}}$ denotes the \stETH to \ETH price in the secondary market. Subsequently, \user supplies \stETH on Aave to borrow $S\cdot p_{t_0}^{a}\cdot l/p_{t_0}^{\text{2nd}}$~\ETH. The borrowed \ETH is swapped again for \stETH within the Curve pool. Although not engaging in direct staking on Lido, \user can still accrue staking rewards, as \stETH employs a rebasing mechanism for reward distribution (see Section~\ref{sec:lsd_intro}).

\item[Leverage Multiplier.] Assume \user invests a total asset (staked \ETH on Lido or swapped \ETH within Curve) of \Asset~\ETH through leverage staking with an initial investment of \principal~\ETH, the leverage multiplier \LEM=$\frac{A_{(S,n)}}{S}$~is defined as the ratio between \Asset~and \principal.

\item[Direct vs Indirect Leverage Staking.] Both direct and indirect leverage staking strategies are designed to amplify staking rewards through a recursive methodology. However, direct leverage staking acquires \stETH from the \LSD primary market, whereas indirect leverage staking obtains \stETH from the secondary market. Notably, indirect leverage staking bypasses the staking process, consequently not contributing to the increase in the total staked \ETH within the network. Utilizing the same principal amount of \ETH, \user generally acquires more \stETH through the indirect strategy, as the \stETH to \ETH price in the secondary market is often below $1$. However, this approach incurs additional costs such as price slippage.

\item[Leverage Staking vs Leverage Borrowing.]  While leverage staking and leverage borrowing both exhibit recursive patterns, they differ for several reasons. First, leverage staking primarily targets \LSDs, whereas leverage borrowing is focused on non-\LSD tokens. Second, leverage staking aims to amplify both staking rewards and the deposit-borrow revenue, while leverage borrowing solely seeks to enhance the deposit-borrow revenue (see Table~\ref{tab:lev_stake_vs_borrow}). Third, they bear different risk sources. In addition to market risk, leverage staking involves the risk of slashing (see Section~\ref{sec: intro_pos}). To summarize, although the traditional leverage borrowing strategy shares some similarities with the leverage staking strategy discussed in this paper, they differ in terms of underlying assets, income sources, and associated risk.
\end{description}


\section{Analytical Study}\label{sec:analytical}

This section conducts an analytical study on the leverage staking strategy. We also offer a generalized formalization encompassing other potential scenarios in Appendix~\ref{appx:generalized_formalization}.

\begin{figure}[t]
    \centering
    \includegraphics[width=0.9\columnwidth]{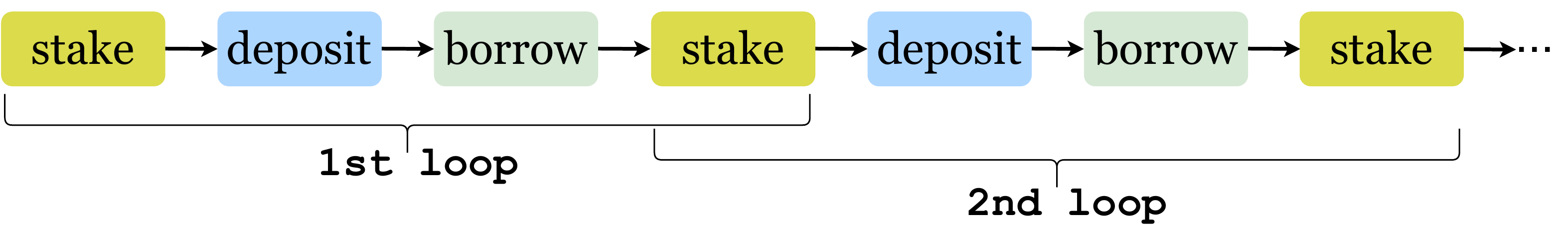}
    \caption{The illustration of direct leverage staking loops. The user completes the $k^{\text{th}}$ loop via a sequence of actions: \texttt{\{stake, deposit, borrow, (re)stake\}}. In parallel, an indirect leverage stake loop is characterized by the sequence \texttt{\{swap, deposit, borrow, (re)swap\}}. Within these frameworks, the (re)stake/(re)swap is crucial in completing the respective loops.}
    \label{fig: lev_loop}
\end{figure}

We assume that \user can complete $n$ loops (see Figure~\ref{fig: lev_loop}) within a short time interval such that the \stETH price remains unchanged. As a rational participant, \user determines the value of $n$ according to its risk profile. Let $p_{t_0}^{a}$ denote the Aave lending price of \stETH and $p_{t_0}^{m}$ be the \stETH to \ETH market price,  where $p_{t_0}^{m}=p_{t_0}^{1st}$ for direct leverage staking and $p_{t_0}^{m}=p_{t_0}^{2nd}$ for indirect leverage staking. \user acquires a total investment amount of \Asset~\ETH, collateral amount of \Collateral~\stETH and debt amount of \Borrow~\ETH (see Equation~\ref{eq:asset_debt}).
\begin{equation}\label{eq:asset_debt} 
   {
    \begin{split}
        & A_{(S,n)} = S\cdot\left[1+ \frac{l\cdot p_{t_0}^{a}}{p_{t_0}^{m}} +...+\left(\frac{l\cdot p_{t_0}^{a}}{p_{t_0}^{m}}\right)^{n}\right]  = S\cdot \frac{1-\left(\frac{l\cdot p_{t_0}^{a}}{p_{t_0}^{m}}\right)^{n+1}}{1-\frac{l\cdot p_{t_0}^{a}}{p_{t_0}^{m}}}\\
        & C_{(S,n)} = \frac{S}{p_{t_0}^{m}}\cdot\left[1+ \frac{l\cdot p_{t_0}^{a}}{p_{t_0}^{m}} +...+\left(\frac{l\cdot p_{t_0}^{a}}{p_{t_0}^{m}}\right)^{n-1}\right] = \frac{S}{p_{t_0}^{m}}\cdot \frac{1-\left(\frac{l\cdot p_{t_0}^{a}}{p_{t_0}^{m}}\right)^{n}}{1-\frac{l\cdot p_{t_0}^{a}}{p_{t_0}^{m}}} \\
        & B_{(S,n)} = S\cdot\left[\frac{l\cdot p_{t_0}^{a}}{p_{t_0}^{m}} +...+\left(\frac{l\cdot p_{t_0}^{a}}{p_{t_0}^{m}}\right)^{n}\right] =S\cdot \frac{\frac{l\cdot p_{t_0}^{a}}{p_{t_0}^{m}}-\left(\frac{l\cdot p_{t_0}^{a}}{p_{t_0}^{m}}\right)^{n+1}}{1-\frac{l\cdot p_{t_0}^{a}}{p_{t_0}^{m}}}
    \end{split} 
    }  
\end{equation}

\begin{figure}[t]
\centering
\begin{minipage}{.495\textwidth}
  \centering
  \includegraphics[width =\columnwidth]
  {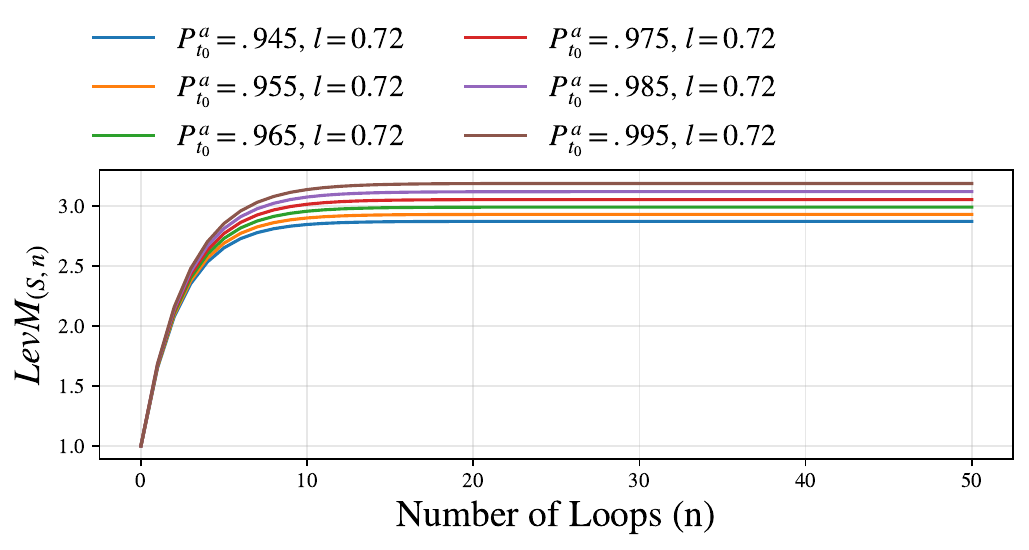}
  \caption{\LEM with varying $ p_{t_0}^{a}$.}
  \label{fig:leverage_Price.pdf}
\end{minipage}%
\hfill
\begin{minipage}{.495\textwidth}
  \centering
  \includegraphics[width =\columnwidth]
  {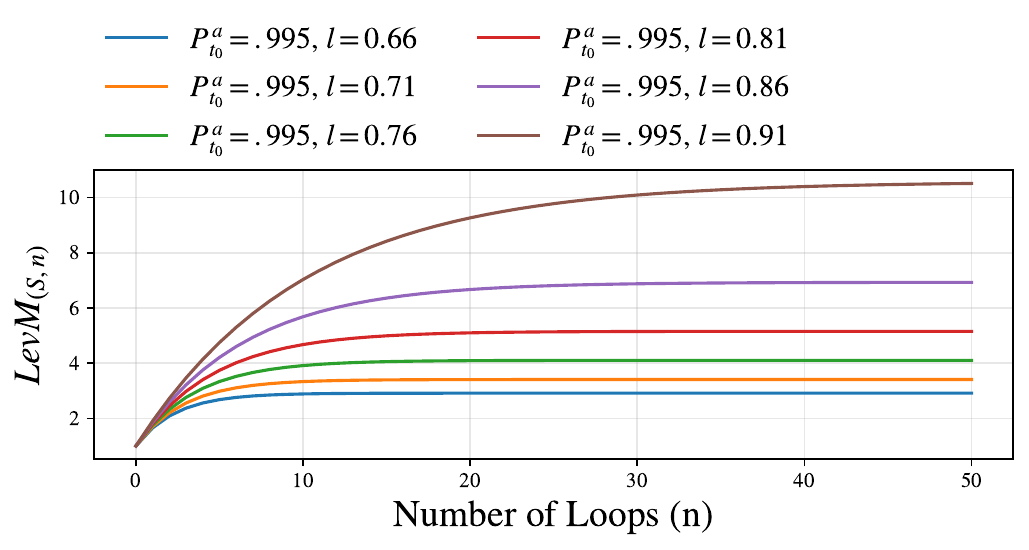}
  \caption{\LEM with varying $l$.}
  \label{fig:leverage_LTV}
\end{minipage}
\end{figure}

\noindent \textbf{Leverage Multiplier.} In Equation~\ref{eq:l_multiplier}, the leverage multiplier is defined as the ratio of \Asset to $S$. For direct leverage staking, we use the primary market price ($p_{t_0}^{m}=p_{t_0}^{1st}=1$). Consequently, \LEM~simplifies to $(1-(l\cdot p_{t_0}^{a})^{n+1})/(1-l\cdot p_{t_0}^{a})$. Figure~\ref{fig:leverage_Price.pdf} and~\ref{fig:leverage_LTV} show how \LEM~changes in response to variations in the \stETH price use by Aave ($p_{t_0}^{a}$) and the \LTVV ratio ($l$) respectively. Notably, when looping towards infinity, the value of $LevM_{(S,n)}$ converges to $1 / (1-l\cdot p_{t_0}^{a})$. For indirect leverage staking where $p_{t_0}^{m}=p_{t_0}^{2nd}$, when $\frac{l\cdot p_{t_0}^{a}}{p_{t_0}^{m}}< 1$, $LevM_{(S,n)}$ coverages towards $1 /(1-\frac{l\cdot p_{t_0}^{a}}{p_{t_0}^{m}})$ as the number of loops approaches infinity.
\begin{equation}\label{eq:l_multiplier}
\centering
{
\begin{split} 
     LevM_{(S,n)} = \frac{A_{(S,n)}}{S} = \frac{1-\left(\frac{l\cdot p_{t_0}^{a}}{p_{t_0}^{m}}\right)^{n+1}}{1-\frac{l\cdot p_{t_0}^{a}}{p_{t_0}^{m}}}  
\end{split} 
 }   
\end{equation}

\noindent \textbf{Health Factor.} Leverage staking can yield high returns but also raises the risk of liquidation. \user's position may be susceptible to liquidation if the value of the collateralized \stETH declines. As discussed in Section~\ref{aave_intro}, Aave uses \HF to track the status of each position. \user's position can be liquidated if \HF$<1$ (see Equation~\ref{health_factor}). In our leverage staking example, the parameters $l$~and \LT are $69\%$ and $81\%$ respectively (see Table~\ref{tab:aave_params} for the changes of Aave parameter configurations). Historically, the Aave v2 always ensures $l<\text{LT}$.

Let $\Delta\% p^{a}_{\Delta t}$ denotes the percentage change of Aave \stETH price from $t_0$ to $t_c$. $\Delta\% p^{a}_{\Delta t}$ is negative if \stETH price decreases. When \stETH price increases, Equation~\ref{health_factor_2} always holds. When \stETH price decreases, Equation~\ref{health_factor_2} suggests that, to uphold a secure position with $\mathit{HF}>1$, the largest acceptable percentage decrease in \stETH price is $\frac{l}{\text{LT}}-1=-\frac{12}{81}\approx -14.8\%$. In a liquidation event, the user's entire collateralized \stETH will be liquidated, and this effect becomes more pronounced as the number of loops ($n$) increases. 

{\footnotesize
\begin{align} 
    & \mathit{HF}_{\mathsf{U_i}}(p_{t_c}^{a}|p_{t_0}^{a}) = \frac{\sum_n^{k} k^{th}\text{ collateralized }\texttt{stETH}\text{ value in }\texttt{ETH}\cdot \text{LT} }{\sum_n^{k} k^{th}\text{ borrowed }\texttt{ETH}\text{ value}} =\frac{C(S,n)\cdot p_{t_c}^{a}\cdot \text{LT}}{B(S,n)} = \frac{p_{t_c}^{a}\cdot \text{LT}}{p_{t_0}^{a}\cdot l} \label{health_factor}\\
    & \mathit{HF}_{\mathsf{U_i}}(p_{t_c}^{a}|p_{t_0}^{a})\geq 1 \implies \frac{p_{t_c}^{a}}{p_{t_0}^{a}}> \frac{l}{\text{LT}} \implies \Delta\% p^{a}_{\Delta t}\geq \frac{l}{\text{LT}}-1 \label{health_factor_2}
\end{align}
}

\noindent \textbf{Profit Breakdown.} 
Equation~\ref{eq:APR} calculates leverage staking profitability. Let $r_s$, $r_c$, $r_b$ represent the staking \APR offered by Lido and the deposit and borrow interest rates provided by Aave respectively. It is worth noting that $r_s$ changes in accordance with the validator performance, while $r_c$ and $r_b$ vary based on Aave's interest rate model\footnote{\url{https://docs.aave.com/risk/liquidity-risk/borrow-interest-rate}}. \user earns a staking \APR of $R_s(n)$ and a deposit \APR of $R_c(n)$, while pays a borrow \APR of $R_b(n)$. In the case of leverage staking, the factor by which $R_s(n)$ is amplified is the total amount of the investment divided by the initial amount of the investment (i.e., $\frac{A_{(S,n)}}{S}$ = \LEM). Similarly, the factor by which $R_c(n)$ is amplified is the total amount of the collateral divided by the initial amount of the collateral (i.e., $\frac{C_{(S,n)}}{S/p_{t_0}^{m}}$). The same logic applies to $R_b(n)$. As such, \user obtains a net APR of $R_{\mathit{Net}}(n)=R_s(n)+R_c(n)-R_b(n)$. The necessary condition for a rational \user to apply leverage staking rather than conventional staking is $R_{\mathit{Net}}(n)>r_s$.
\begin{equation}\label{eq:APR}
   {
   \footnotesize
    \begin{split}
        R_{\mathit{Net}}(n) = R_s(n)+R_c(n)-R_b(n) = r_s \cdot\frac{A_{(S,n)}}{S}+r_c \cdot\frac{C_{(S,n)}\cdot p_{t_0}^{m}}{S}-r_b \cdot\frac{B_{(S,n)}\cdot p_{t_0}^{m}}{S\cdot l\cdot p_{t_0}^{a}}
    \end{split}
    }
\end{equation}

In addition to the standardized scenario discussed above, real-world applications of leverage staking can vary significantly among users. For instance, a user might choose not to reinvest all of their received \stETH on Aave. For a more detailed exploration of this variability, please see the generalized formalization in Appendix~\ref{appx:generalized_formalization}.


\section{Empirical Study} \label{sec: empirical}
We outline the empirical evaluation of leverage staking across Aave, Lido and Curve.


\smallskip \noindent\textbf{Data Collection.} We first crawl the on-chain events involving users' actions on the Aave V$2$ lending pool, including \texttt{deposit}, \texttt{borrow}, \texttt{withdraw}, and \texttt{repay} events. For direct leverage staking, we crawl the historical \texttt{stake} (i.e, \texttt{submitted}) events related to Lido \href{https://etherscan.io/address/0xae7ab96520de3a18e5e111b5eaab095312d7fe84}{\stETH} token when users stake \ETH on Lido. For indirect leverage staking, we crawl the historical \texttt{swap} (i.e., \texttt{TokenExchange}) events for Curve \href{https://etherscan.io/address/0xdc24316b9ae028f1497c275eb9192a3ea0f67022}{\stETH--\ETH} pool. We use an Ethereum Geth node on a Linux machine running Ubuntu 22.04 LTS, which is equipped with AMD $48$-core CPU, $256$ GB of RAM, and $12 \times 2$ TB SSD. We capture all the targeted events from block $11{,}473{,}216$ (Dec~$17$,~$2020$) to block $17{,}866{,}191$ (Aug~$7$,~$2023$), $963$ days in total. We identify $290{,}984$ \texttt{stake} events on Lido, $449{,}528$ \texttt{deposit}, $238{,}388$ \texttt{borrow}, $336{,}746$ \texttt{withdraw} and $173{,}596$ \texttt{repay} events on Aave V$2$ lending pool, and $105{,}310$ \texttt{swap} events on Curve \stETH--\ETH pool.

\smallskip \noindent\textbf{Leverage Staking Detection.} We proceed to analyze the users who adopt the direct or indirect leverage staking strategy. From the $449{,}528$ \texttt{deposit} and $238{,}388$ \texttt{borrow} events on Aave V$2$, we find that $743$ addresses are used to deposit \stETH as collateral and then borrow \ETH. We then propose Algorithm~\ref{alg: leverage-staking-detection} and \ref{alg: indirect-leverage-staking-detection} (see Appendix~\ref{appx:algo}) to identify the addresses involving direct and indirect leverage staking respectively. Specifically, we extract the event sequences of $(\texttt{stake}$, $\texttt{deposit}$, $\texttt{borrow}$, $\texttt{stake})$ and $(\texttt{swap}$, $\texttt{deposit}$, $\texttt{borrow}$, $\texttt{swap})$  in chronological order, which follows the direct and indirect leverage staking process (Figure~\ref{fig: lev_loop}).

\begin{figure}[t]
\centering
\begin{minipage}{.495\textwidth}
  \centering
  \includegraphics[width =\columnwidth]
  {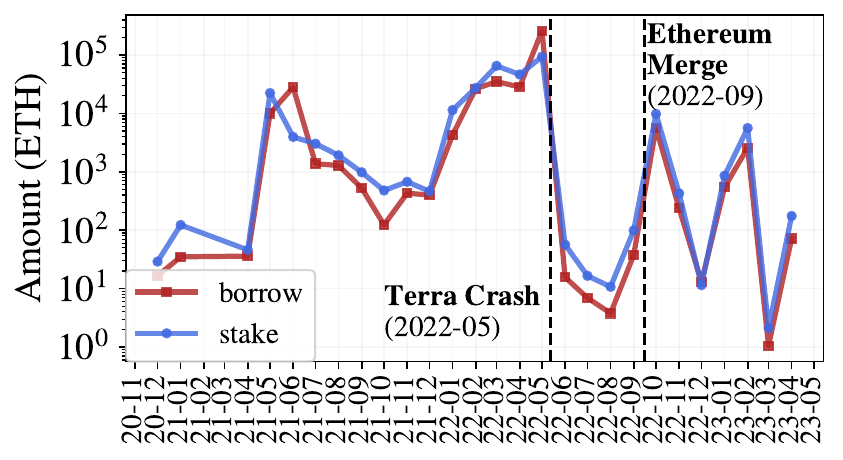}
  \caption{Direct leverage staking statistics.}
  \label{fig:lido_revolving_loan_amount_over_time}
\end{minipage}%
\hfill
\begin{minipage}{.495\textwidth}
  \centering
  \includegraphics[width =\columnwidth]
  {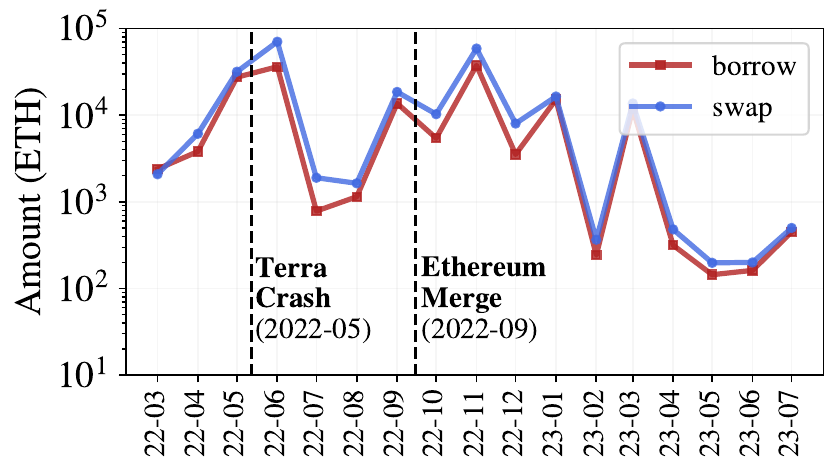}
  \caption{Indirect leverage staking statistics.}
  \label{fig:lido_revolving_loan_amount_over_time_indirect}
\end{minipage}%
\end{figure}

\begin{figure}[t]
\centering
\begin{minipage}{.495\textwidth}
\centering
  \includegraphics[width =\columnwidth]
  {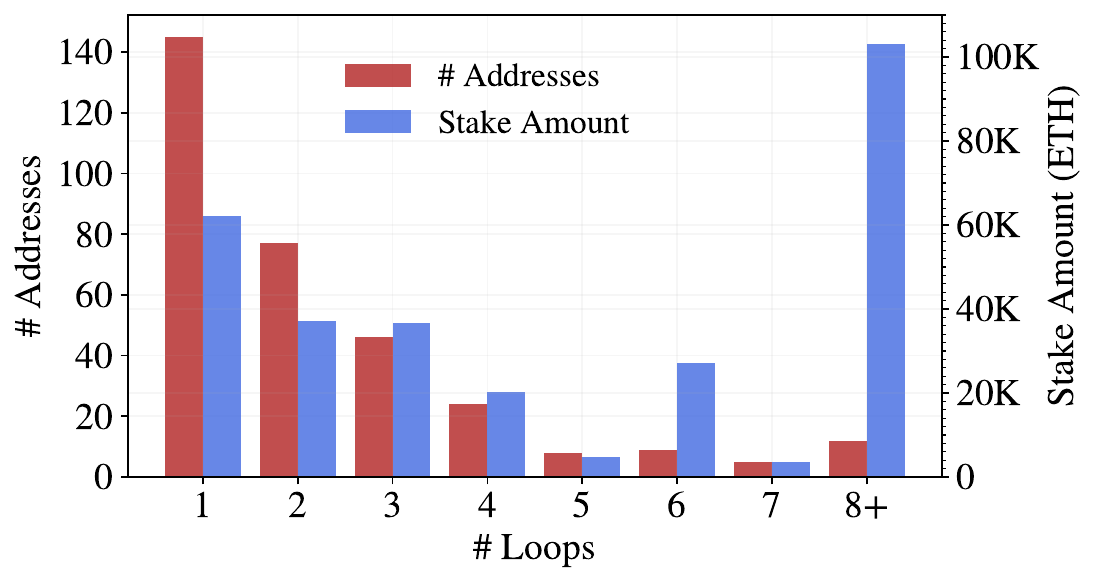}
  \caption{Stake amount and the number of direct leverage staking addresses by loops.}
  \label{fig:lido_revolving_loan_address_distribution}
\end{minipage}
\hfill
\begin{minipage}{.495\textwidth}
\centering
  \includegraphics[width =\columnwidth]
  {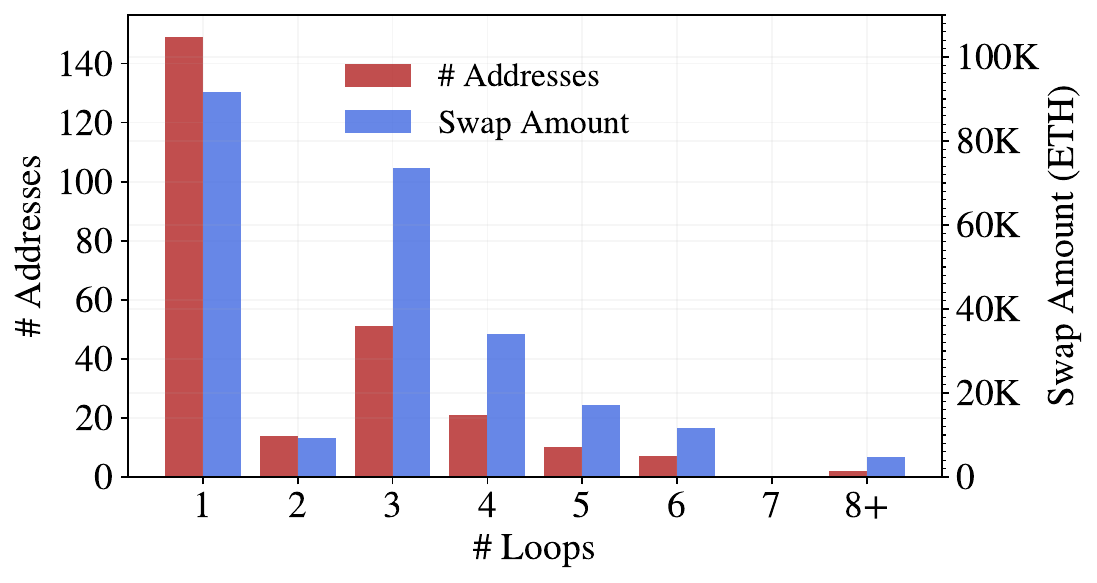}
  \caption{Swap amount and the number of indirect leverage staking addresses by loops.}
  \label{fig:lido_revolving_loan_address_distribution_indirect}
\end{minipage}
\end{figure}

We have identified 262 addresses that have been engaging in direct leverage staking activities, with a cumulative stake amount of \totalLeverageStakingAmt~\ETH. In addition, we observe $180$ addresses that have performed the indirect leverage staking strategy, with a cumulative swap amount of $241{,}880$~\ETH. The distribution of leverage stake and swap amount is depicted in Figure~\ref{fig:lido_revolving_loan_amount_over_time} and \ref{fig:lido_revolving_loan_amount_over_time_indirect} respectively.
Interestingly, we observe that the volume of both direct and indirect leverage staking was substantially impacted by the Terra crash incident in May~$2022$. The stake amount of direct leverage staking experienced a drastic decline from the peak monthly stake amount of $93{,}661$~\ETH in May~$2022$ to $11$~\ETH in Aug~$2022$. Similarly, the swap amount of indirect leverage staking declines from $70{,}655$~\ETH in Jun~$2022$ to $1{,}639$~\ETH in Aug~$2022$. Later, the Ethereum Merge brought about a resurgence in leverage staking activities, with a stake amount of $9{,}814$~\ETH and a swap amount of $10{,}293$~\ETH in Nov~$2022$.

\smallskip \noindent\textbf{Leverage Staking Loops.} Among $262$ and $180$ addresses that have adopted direct and indirect leverage staking, we conduct an analysis focusing on two key elements: the number of loops (denoted as $n$) and the leverage multiplier (denoted as \LEM), derived from their extracted action sequence $\mathcal{E}_s$ (see~Algorithm~\ref{alg: leverage-staking-detection} and~\ref{alg: indirect-leverage-staking-detection}). To calculate the number of direct and indirect leverage staking loops, we identify consecutive sub-sequences in $\mathcal{E}_s$ consisting of ($\texttt{stake}$, $\texttt{deposit}$, $\texttt{borrow}$, $\texttt{stake}$) and ($\texttt{swap}$, $\texttt{deposit}$, $\texttt{borrow}$, $\texttt{swap}$) respectively. Figure~\ref{fig:lido_revolving_loan_address_distribution} reveals that $145$ addresses ($55.35\%$) performed direct leverage staking with a single loop ($n=1$), while Figure~\ref{fig:lido_revolving_loan_address_distribution_indirect} shows that $149$ addresses ($82.78\%$) performed indirect leverage staking with a single loop. Although only a smaller subset of $12$ addresses performs direct leverage staking with more than eight loops, their cumulative staking activities amount to a significant volume of $102{,}998$~\ETH.  This highlights a concentrated yet substantial engagement in direct leverage staking. In contrast, only $2$ addresses performed indirect leverage staking with more than eight loops, with a total swap amount of $4{,}669$~\ETH. The difference in the number of participants and the total amount staked suggests distinct participant profiles and strategies between direct and indirect leverage staking. The comparison indicates that direct leverage staking is more likely to attract sophisticated market participants who are willing to perform more loops with substantial capital commitments.

\begin{figure}[t]
\centering
\begin{minipage}{.495\textwidth}
  \centering
  \includegraphics[width =\columnwidth]
  {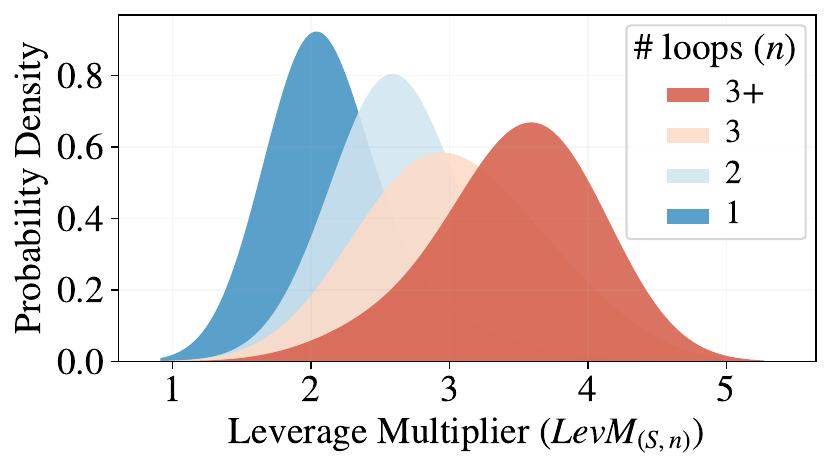}
  \caption{Distribution of direct leverage staking \LEM~by loops.}
  \label{fig:lido_LevM_density}
\end{minipage}%
\hfill
\begin{minipage}{.495\textwidth}
  \centering
  \includegraphics[width =\columnwidth]
  {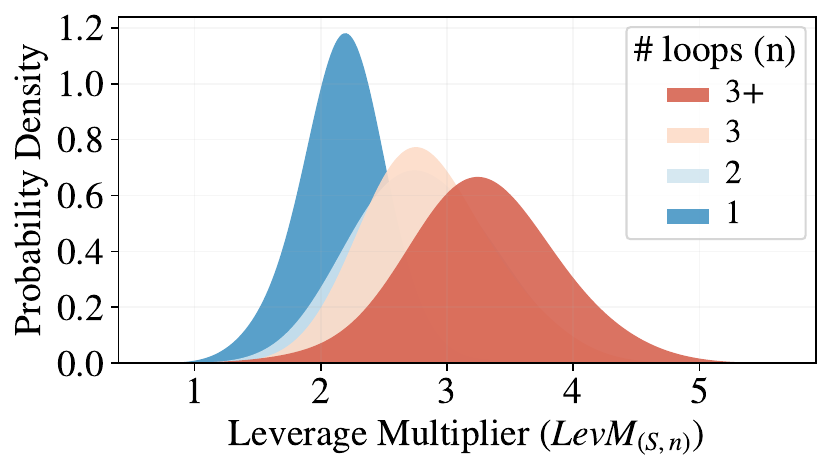}
  \caption{Distribution of indirect leverage staking \LEM~by loops.}
  \label{fig:lido_LevM_density_indirect}
\end{minipage}
\end{figure}

\begin{figure}[t]
\centering
\begin{minipage}{.495\textwidth}
  \centering
  \includegraphics[width =\columnwidth]
  {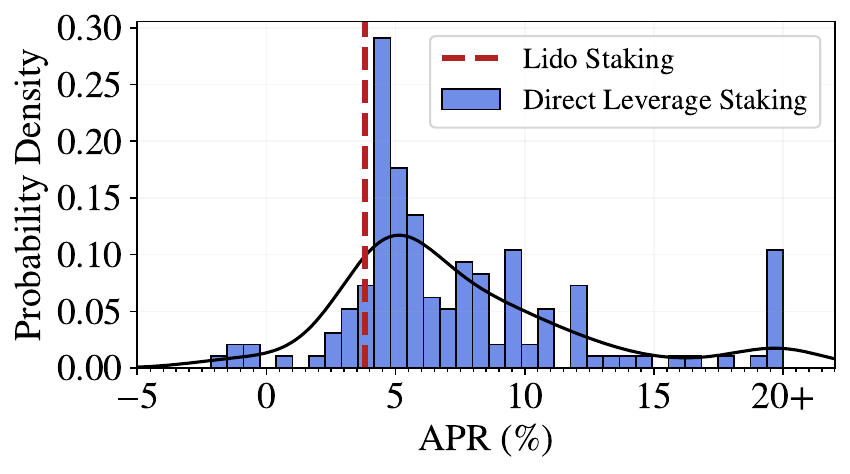}
  \caption{Direct leverage staking APR.}
  \label{fig:lido_apr_distribution}
\end{minipage}
\hfill
\begin{minipage}{.495\textwidth}
  \centering
  \includegraphics[width =\columnwidth]
  {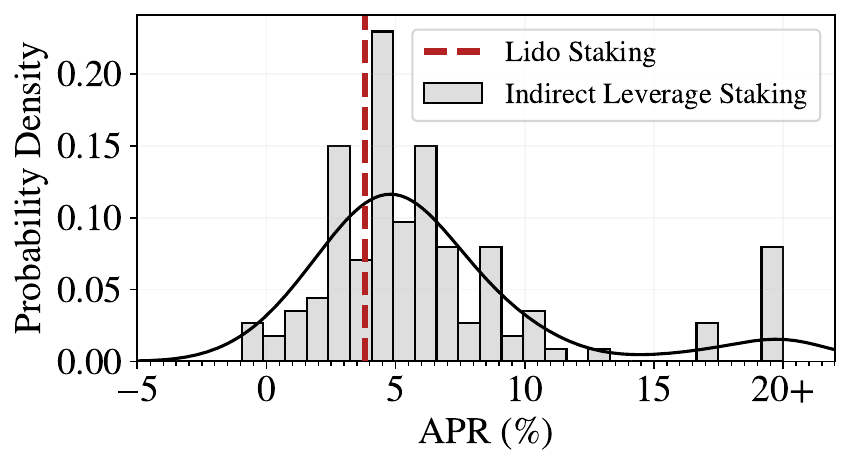}
  \caption{{Indirect leverage staking APR.}}
  \label{fig:lido_apr_distribution_indirect}
\end{minipage}
\end{figure}

\smallskip \noindent\textbf{Leverage Staking Multipliers.} Furthermore, we compute \LEM~for each address by taking into account the initial and the cumulative sum of \texttt{stake} or \texttt{swap} amount~(See~Equation~\ref{eq:l_multiplier}). Figure~\ref{fig:lido_LevM_density} and~\ref{fig:lido_LevM_density_indirect} illustrate the distribution of \LEM~across various $n$. The trend shows that a higher loop count $n$ generally corresponds to higher \LEM~in practical scenarios. Additionally, it is noteworthy that the majority (more than $90\%$) of the direct and indirect leverage staking addresses exhibit a \LEM~smaller than $4$.

\smallskip \noindent\textbf{Leverage Staking APR.} We focus on a subset of $152$ and $137$ direct and indirect leverage staking addresses that have successfully repaid their debts and withdrawn their collateral from Aave \stETH--\ETH positions. 
To calculate their \emph{actual} \APR, as outlined in Equation~\ref{eq: actualAPR}, we consider the net earnings from \texttt{deposit} and \texttt{withdraw} actions, balanced against the \ETH accrued through \texttt{borrow} and \texttt{repay} actions. Additionally, we account for the conversion of accrued \ETH to \stETH, factoring in the \stETH price at the time of the last \texttt{withdraw} action. 
\begin{equation}\label{eq: actualAPR}
\begin{aligned}
    &\mathsf{actualAPR} = \frac{\left(\mathsf{accruedstETH} - \frac{\mathsf{accuredETH}}{p_{\stETH}}\right)\cdot \frac{3600\cdot 24 \cdot 365}{12}}{\mathsf{totalDepositstETH} \cdot \left(\mathsf{lastWithdrawBlock} - \mathsf{firstDepositBlock}\right)}\\
    &\mathsf{accruedstETH} = \mathsf{totalWithdrawstETH}-\mathsf{totalDepositstETH} \\
    &\mathsf{accruedETH} = \mathsf{totalRepayETH}-\mathsf{totalBorrowETH}\\
\end{aligned}
\end{equation}

The distributions of direct and indirect leverage staking \APR are visually depicted in Figure~\ref{fig:lido_apr_distribution} and~\ref{fig:lido_apr_distribution_indirect} respectively. Notably, our findings reveal that a significant majority ($81.7\%$), precisely $137$ ($90.13\%$) direct and $99$ ($73.33\%$) indirect leverage staking addresses have realized an \APR higher than the \APR of conventional staking on Lido, highlighting the financial opportunities of such strategy.

\begin{figure}[t]
\centering
\begin{minipage}{.49\textwidth}
  \centering
    \includegraphics[width=0.95\columnwidth]{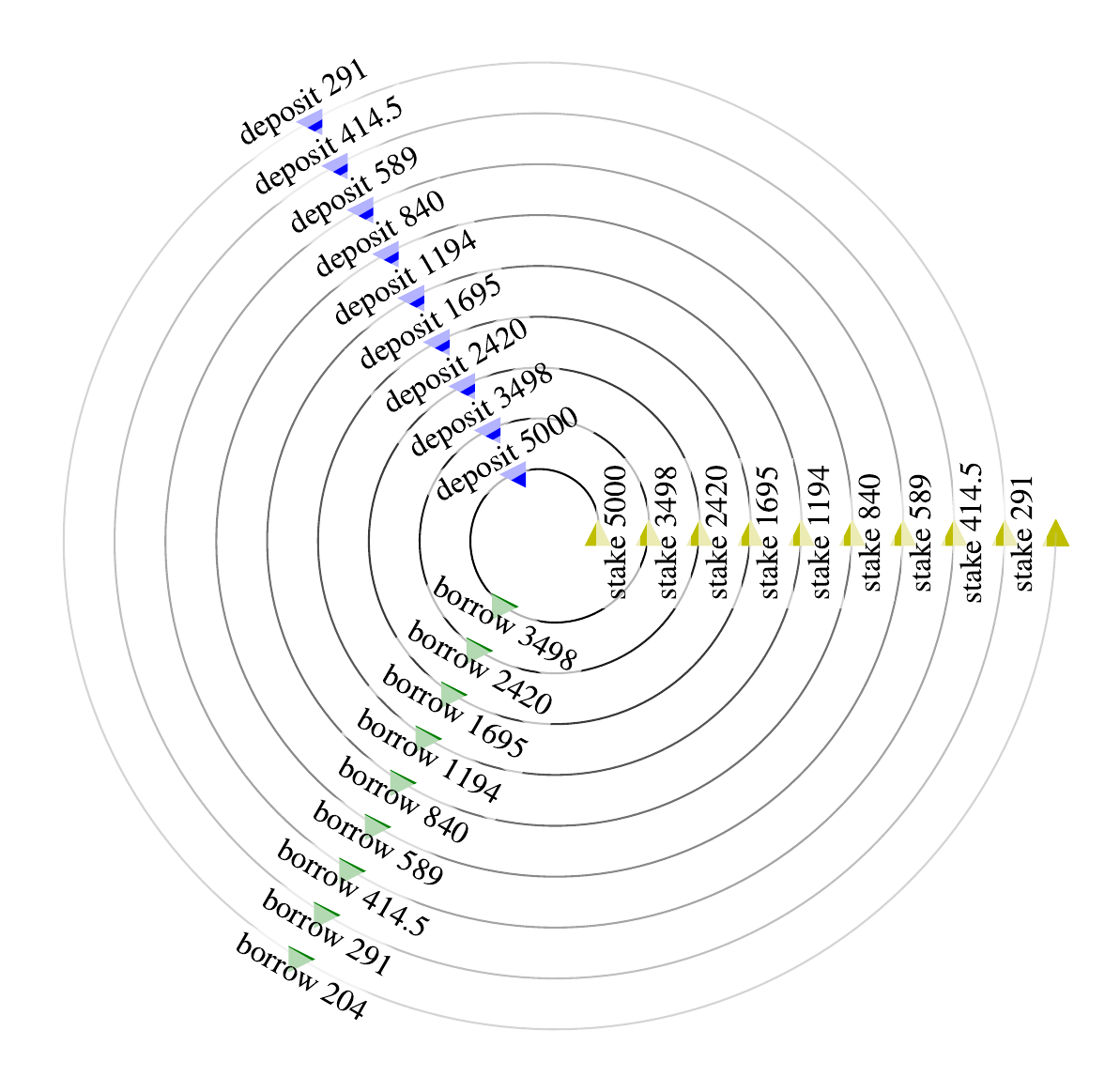}
    \caption{\oxd: Direct leverage staking from block $14{,}617{,}906$ to $14{,}627{,}202$.}
    \label{fig: case_study_1}
\end{minipage}%
\hfill
\begin{minipage}{.49\textwidth}
  \centering
    \includegraphics[width=0.95\columnwidth]{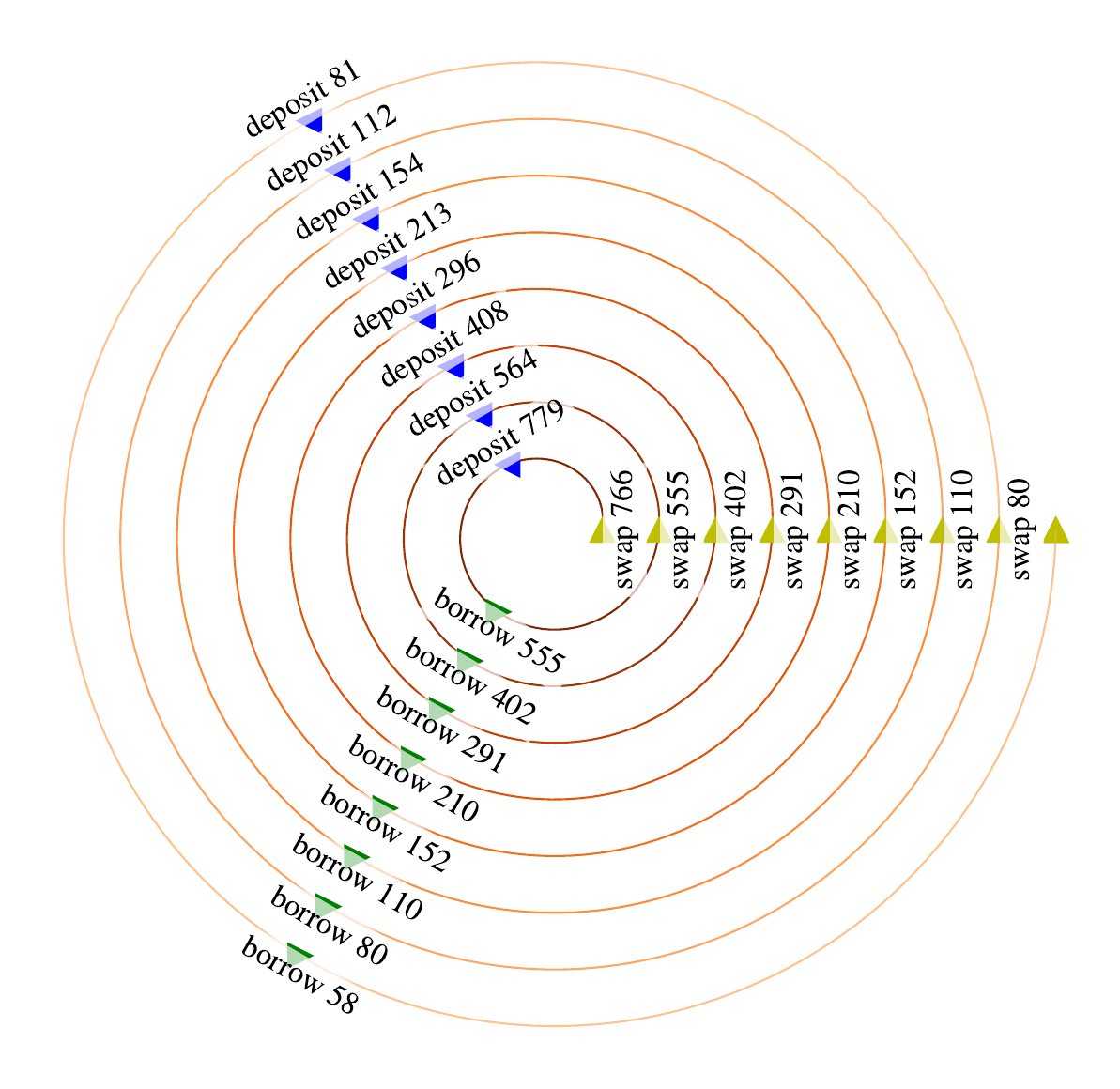}
    \caption{\oxA: Indirect leverage staking from block $16{,}031{,}087$ to $16{,}031{,}208$.}
    \label{fig: case_study_2}
\end{minipage}
\end{figure}

\smallskip
\noindent\textbf{Leverage Staking Examples.} We present two examples to enhance the understanding of leverage staking. From block $14{,}617{,}906$ to $14{,}627{,}202$, a whale wallet address \oxd executed the direct leverage staking strategy with $9$ loops. The recursive action sequences of (\texttt{stake, deposit, borrow, stake}) are shown in Figure~\ref{fig: case_study_1}. \oxd invested a principal amount of $5{,}000$ \ETH. The direct leverage staking results in a total investment amount of $16{,}145.5$ \ETH. This led to a leverage multiplier of $3.23$, demonstrating the amplification effect of the leverage strategy. In another instance, from block $16{,}031{,}087$ to $16{,}031{,}208$, \oxA performed the indirect leverage staking strategy. This was accomplished by recursively executing the action sequence of (\texttt{swap, deposit, borrow, swap}). \oxA started with a principal investment amount of $766$ \ETH. Through $8$ leverage loops, \oxA achieved a total investment of $2{,}624$ \ETH, resulting in a leverage multiplier of $3.43$. This illustrates the effectiveness of the indirect leverage staking strategy in increasing the total investment.


\section{Cascading Liquidation}

In this section, we offer an overview of the \stETH price deviation in the context of the Terra crash incident. We illustrate how the \stETH price can potentially lead to liquidation cascades within the \LSD ecosystem,  particularly in the context of leverage staking.

\begin{figure}[tbh]
    \centering
    \includegraphics[width=0.95\columnwidth]{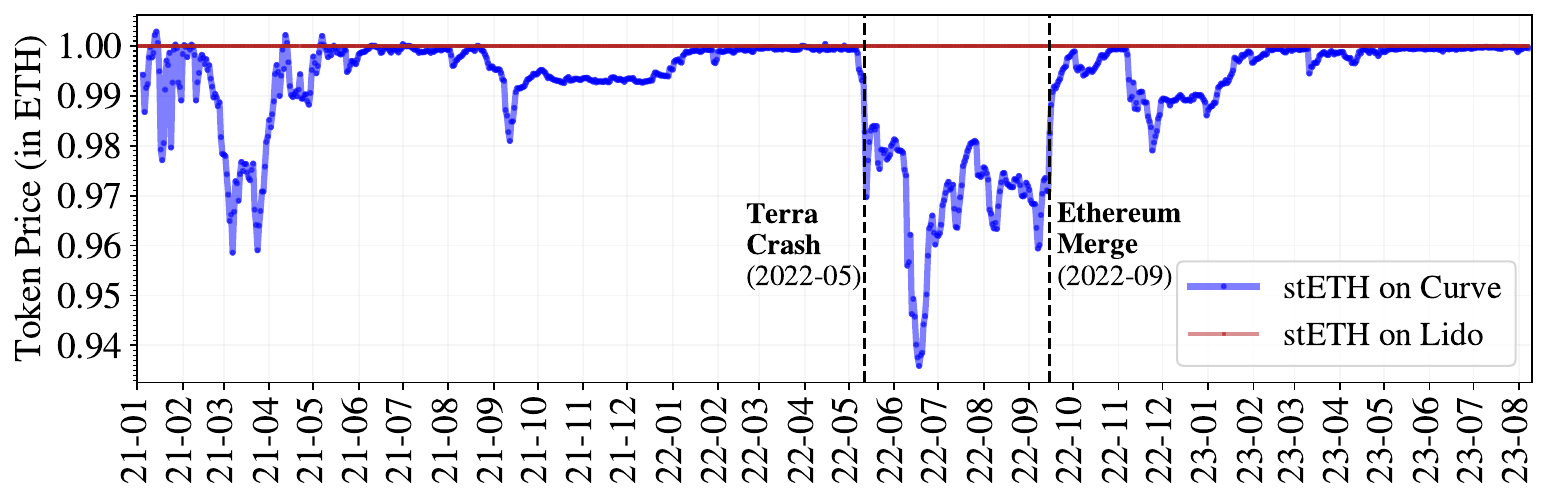}
    \caption{\stETH price in the primary (Lido) and secondary (Curve) markets.}
    \label{fig: lido_wstETH_price_over_time}
\end{figure}

\begin{figure}[tbh]
    \centering
    \includegraphics[width=\columnwidth]{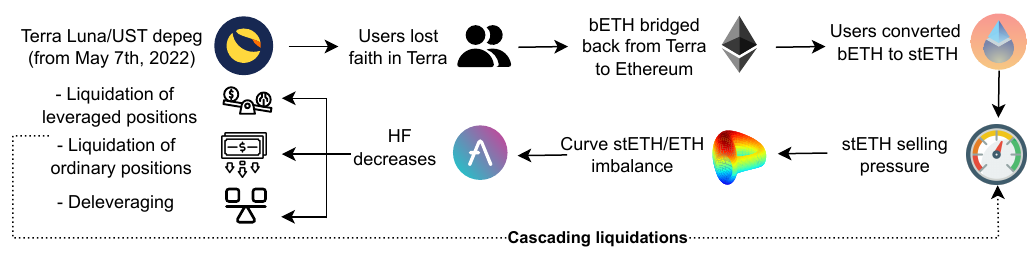}
    \caption{Illustration of liquidation cascades.}
    \label{fig: cascad_liqs}
\end{figure}

\subsection{stETH Price Deviation and Terra Crash Incident}
As a rebasing \LSD, \stETH changes its token supply to distribute rewards to stakers (see Section~\ref{sec:lsd_intro}). As such, the \stETH to \ETH price in the primary market (i.e., Lido) is $1$. While \stETH is not required to trade on par with \ETH in the secondary market (e.g., Curve), the price is anticipated to converge to $1$. Our empirical data show that \stETH did maintain a loose peg to \ETH for most of its history. However, the \stETH price began to drop from May~$12$,~$2022$, reaching its lowest point of $0.931$ on May~$18$,~$2022$~(see Figure~\ref{fig: lido_wstETH_price_over_time}). 

The \stETH price decline can date back to the \UST/\LUNA depeg. The Terra collapse instilled fear and triggered selling pressure throughout the market~\cite{liu2023anatomy,lee2023dissecting}. Specifically, following the \UST/\LUNA depeg incident between May $7$ to $16$, investors grew concerned about the security and stability of the Terra network. Given the prevailing bearish sentiment, investors moved to bridge back \bETH (a wrapped version of \stETH on Terra) from Terra to Ethereum via the \href{https://etherscan.io/address/0x3ee18b2214aff97000d974cf647e7c347e8fa585}{{Wormhole contract}}. Our data show that $614$k \bETH was bridged to Ethereum, with a remarkable $98\%$ of these \bETH converted back to \stETH. This mass conversion reflects the widespread desire to exit Terra-based staking assets. Subsequently, the secondary market experienced significant selling pressure, primarily from institutions such as \href{https://celsius.network/}{Celsius}. This imbalance in the Curve \stETH--\ETH pool contributed to the further decline of \stETH price.

\subsection{Cascading Liquidation and User Behaviors}\label{sec:liquidation-risk}

The decline in \stETH price may trigger liquidation cascades within the \LSD ecosystem, especially in the context of leverage staking~(see Figure~\ref{fig: cascad_liqs}). Specifically, the decline in \stETH price reduces the \HFs of \stETH collateralized borrowing positions on Aave, potentially leading to liquidations. In response to liquidations, users with leverage staking positions can either take no action or choose to deleverage their positions.

On the one hand, users with leverage staking positions may take no action when their \HFs approach the critical threshold of $1$. In this case, their collateralized \stETH might be liquidated. The liquidators repay \ETH to acquire \stETH, with the liquidation amount being amplified by \LEM. Subsequently, a significant amount of \stETH is sold in the Curve pool, contributing to additional selling pressure on \stETH (see Figure~\ref{fig: cascad_liqs}). This extensive selling further imbalances the Curve \stETH--\ETH pool, resulting in a further decline in \stETH price. Consequently, an increasing number of positions, including both leverage staking and ordinary positions, are vulnerable to liquidation as a result of declining \HFs.

On the other hand, users can choose to deleverage their positions and restore \HFs. Assuming \user has executed a direct or indirect leverage staking strategy with $n$ loops, \user can initiate a deleveraging process with the following steps. \emph{(i)} \user executes a swap to convert \stETH into \ETH within the Curve \stETH--\ETH pool. \emph{(ii)} The received \ETH is then used to repay the \ETH borrowed in the $n^\text{th}$ loop. \emph{(iii)} \user then withdraws the \stETH that was supplied in the $n^\text{th}$ loop from Aave and continues converting it into \ETH using the Curve pool. This ``\texttt{swap-repay-withdraw}'' process is repeated as necessary to restore \HFs.

\begin{figure}[t]
\centering
\begin{minipage}{.495\textwidth}
  \centering
  \includegraphics[width =\columnwidth]
  {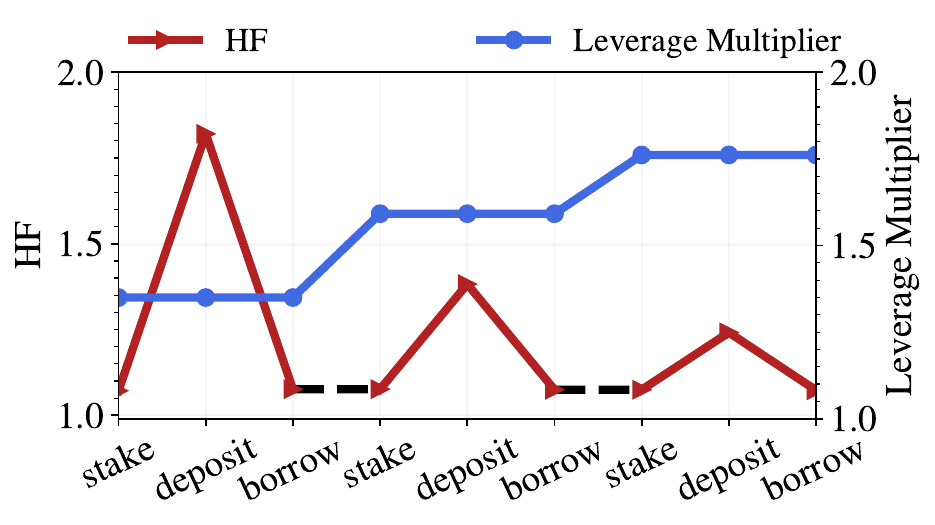}
  \caption{Example of the leverage action.}
  \label{fig:lido_leverage_HF}
\end{minipage}%
\hfill
\begin{minipage}{.48\textwidth}
  \centering
  \includegraphics[width =\columnwidth]
  {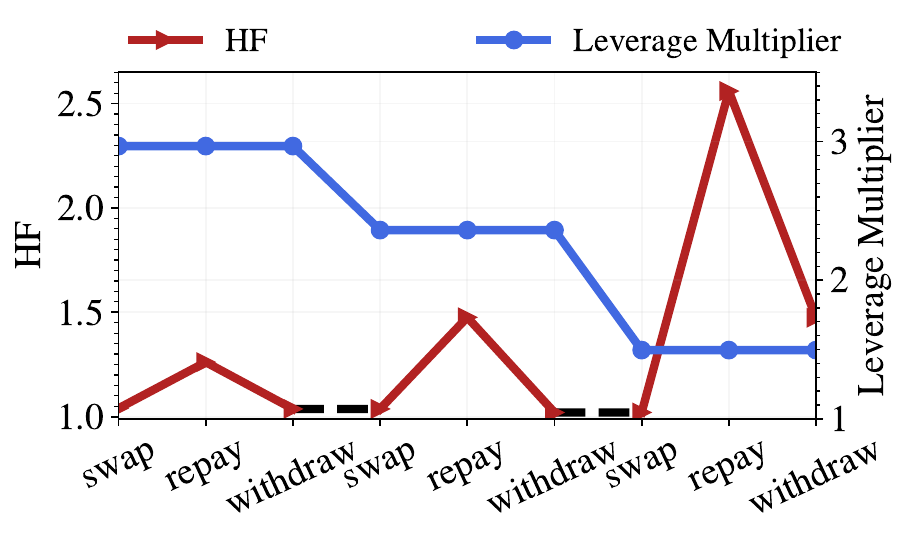}
  \caption{Example of the deleverage action.}
  \label{fig:lido_deleverage_HF}
\end{minipage}
\end{figure}

Taking address \oxd as an example, the overall trend of \HF and \LEM~during the deleveraging process (see Figure~\ref{fig:lido_deleverage_HF}) exhibit a remarkable degree of symmetry when compared to those observed in the leveraging process (Figure~\ref{fig:lido_leverage_HF}). With each \texttt{repay} action, the \HF of the address increases, as indicated by the red line, while the leverage multiplier decreases, as shown by the blue line. This symmetrical trend illustrates the correlation between repaying debt and improving \HF, which consequently reduces \LEM.

During the period from \TerraCrashStartDate to \TerraCrashStartDateAndTen, i.e., the first ten days after the Terra crash (Figure~\ref{fig: lido_wstETH_price_over_time}), we observed $13$ users actively deleveraging their direct leverage staking positions and $5$ users deleveraging their indirect leverage staking positions. This activity resulted in a total debt repayment of $74{,}983.6$~\ETH and $61{,}085.5$~\ETH respectively. However, it is important to note that even if a user manages to avoid liquidation by deleveraging, the additional selling pressure generated by the \texttt{swap} transactions on Curve can still intensify the decline in \stETH price. This decline may increase the vulnerability of other leverage staking and ordinary positions to liquidation. Such contagion effect may increase market instability and affect a broader range of participants beyond those directly engaged in deleveraging.

\smallskip
To summarize, users with leverage staking positions can take various actions to respond to potential liquidations. Regardless of their choices, these actions may contribute to additional selling pressure on \stETH, further exacerbating price declines and liquidation cascades. This dynamic underscores the interconnection of leverage staking and the broader market ecosystem. In the following section, we will conduct stress tests to evaluate such risks.

\section{Stress Testing}
\subsection{Motivation}

By crawling the \texttt{liquidationcall} events on Aave V$2$ lending pool from \LidoStartCrawlingDate to \LidoEndCrawlingDate, we identify $18$ liquidations for the positions where users supplied \stETH to borrow \ETH, $7$ liquidations for direct leverage staking positions, and $2$ liquidations for indirect leverage staking positions. This relatively low number of liquidations can be attributed to the fact that \stETH has historically only experienced a modest price decline, with a low of $0.931$. However, drawing from the \LUNA--\UST incident, we recognize that a token may become entirely devalued. Should \stETH undergo a devaluation similar to that of \LUNA, it could trigger a surge in liquidations. Therefore, it is crucial to conduct stress tests to assess the risk of cascading liquidations under the worst-case scenario.

\subsection{Simulation}

Motivated by these concerns, we perform stress tests on the Lido-Aave-Curve LSD ecosystem under extreme conditions, simulating potential liquidation events, selling pressures, and subsequent liquidation cascades triggered by a significant \stETH devaluation. Our objective is to address the following simulation questions (SQs):
\begin{description}
    \item[SQ1] How are leverage staking positions affected by \stETH devaluation?
    \item[SQ2] How does the liquidation of leverage staking positions affect the price of \stETH?
    \item[SQ3] How do leverage staking positions affect ordinary positions during \stETH devaluation?
    \item[SQ4] What are the effects of deleveraging actions on \stETH price and market participants?
\end{description}

We first categorize Aave collateralized \stETH borrowing positions into two groups: the leverage staking group (\GroupLV) and the ordinary group (\GroupOD). We then simulate four distinct scenarios to investigate the answers to the corresponding SQs.

\subsubsection{SQ1}
SQ1 aims to simulate the impact of a significant \stETH devaluation on leverage staking positions. We focus on the fluctuations in \HFs and record the number of liquidated positions.

\smallskip
\noindent\emph{Simulation Setup.} We initialize the Curve \href{https://etherscan.io/address/0xDC24316b9AE028F1497c275EB9192a3Ea0f67022#code}{\stETH--\ETH pool} by forking its state at block $17{,}500{,}000$ (Jun~$17$,~$2023$), with reserve of $265{,}972$ \ETH and $266{,}966$ \stETH. Subsequently, we mimic the institutional selling pressure (e.g.,  Celsius; see Figure~\ref{fig: cascad_liqs}) after the Terra crash incident by simulating a sale of $170{,}000$ \stETH on Curve. This sizable transaction causes a decline in the \stETH price, resulting in a new exchange rate of $100~\stETH = 90.52~\ETH$, denoted as $p_0 = 0.9052$. In addition, we initialize \GroupLV with $262$ direct and $180$ indirect leverage staking positions, each with an \texttt{address} that we have detected in Section~\ref{sec: empirical}. For each position, the values of \texttt{totalDebtETH}, \texttt{totalCollateralETH}, and \texttt{HF} are set to the corresponding values recorded in the transaction logs of that position's most recent borrowing transaction. Furthermore, the \texttt{stETHPrice} for all positions is initialized as $p_0$.

\smallskip
\noindent\emph{Simulation Process.} The simulation process consists of a series of sequential rounds. In each round, the \texttt{stETHPrice} for all positions is updated as the current \stETH price in the Curve \stETH--\ETH pool. Subsequently, the \HF for each position is recalculated, using the updated \texttt{stETHPrice}. If, at any point, a position's \HF drops below the threshold of $\mathit{HF}=1$, a simulated liquidation event is triggered. In this scenario, a designated liquidator steps in to settle the debt by repaying it in \ETH. In return, the liquidator receives the collateral in \stETH. All received \stETH is converted to \ETH in the Curve \stETH--\ETH pool, as shown in Figure~\ref{fig: cascad_liqs}. This process continues until no more liquidatable positions remain.

\begin{figure}[tbh]
\centering
\begin{minipage}{.495\textwidth}
  \centering
   \includegraphics[width =\columnwidth]
  {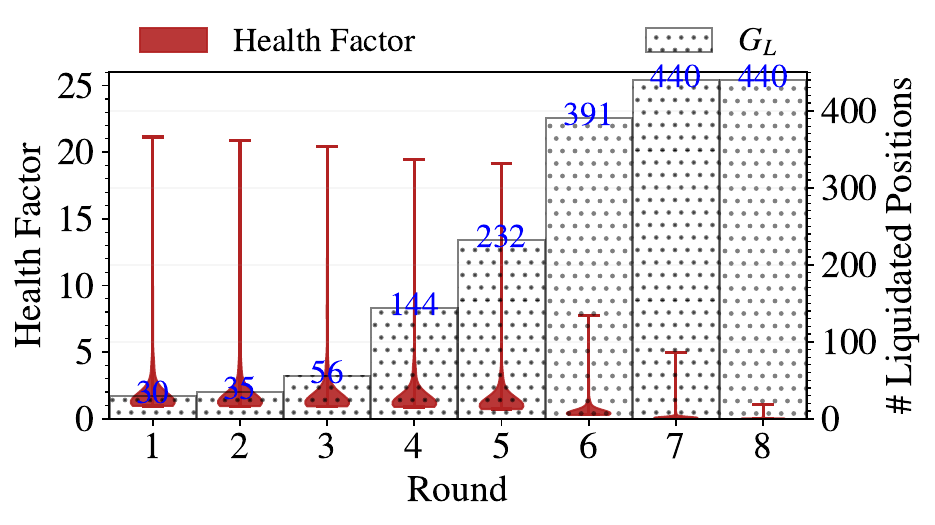}
  \caption{\#~liquidated leverage staking positions and the change of HFs. 
  }
  \label{fig:cascading_liquidation_hf_count}
\end{minipage}%
\hfill
\begin{minipage}{.495\textwidth}
  \centering
  \includegraphics[width =0.95\columnwidth]
  {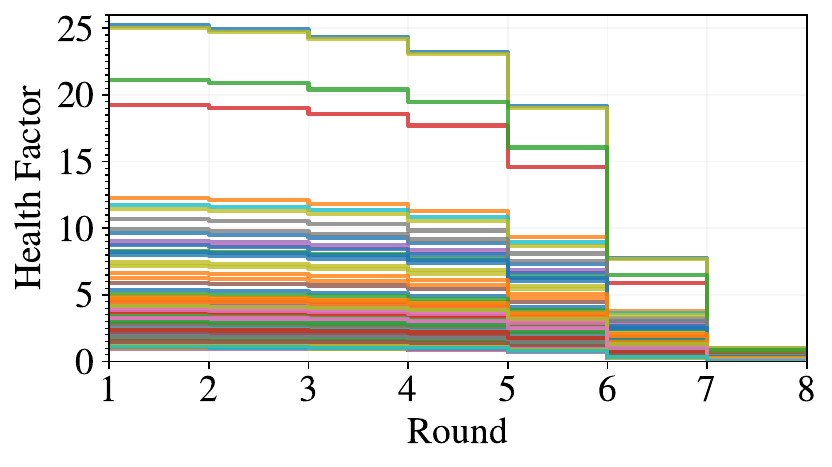}
  \caption{Simulated change in HFs for leverage staking positions in \GroupLV.}
  \label{fig:cascading_liquidation_hf_results}
\end{minipage}
\end{figure}

\smallskip
\insightbox { \label{insights-1}
\normalfont 
During the devaluation of \stETH, leverage staking positions become vulnerable to liquidations due to significant declines in their HFs. 
}

\noindent\emph{Simulation Result.} We examine the liquidation dynamics of \GroupLV's leverage staking positions and the fluctuations in their \HFs throughout simulations. Figure~\ref{fig:cascading_liquidation_hf_count} shows the number of liquidated users in group \GroupLV and the variations in \HFs across different simulation rounds. The simulation terminates after $8$ rounds, where $440\thinspace(99.55\%)$ positions are liquidated. A noteworthy observation is that \HFs of all positions exhibit a steep decrease, as depicted in Figure~\ref{fig:cascading_liquidation_hf_results}. The liquidation cascades experienced by leverage staking positions result in a total liquidated amount $497{,}375$~\ETH, ultimately driving the \stETH price down to $0.01$~\ETH.

\subsubsection{SQ2}
SQ2 aims to explore how the liquidation of leverage staking positions can impact the price of \stETH. We first simulate a scenario to evaluate the effects of leverage staking strategies (\GroupLV), followed by a contrasting scenario where users do not adopt these strategies. The selling pressure originates from the liquidation of the corresponding positions. 

We first simulate a scenario to assess the impact of leverage staking on \stETH price and the liquidation volume. Next, we simulate an alternative scenario in which users within group $G_L$, which includes $262$ direct and $180$ indirect leverage staking positions, do not adopt leverage staking strategies. This involves setting the initial values for \texttt{totalCollateralETH}, \texttt{totalDebtETH}, and \texttt{HF} as the values recorded in the transaction logs when the \emph{first} borrowing action for the position occurred. 

\insightbox { \label{insights-2}
\normalfont 
Leverage staking amplifies the risks of cascading liquidations. The liquidation of leverage staking positions introduces additional selling pressure to the market, thereby exacerbating the decline in \stETH prices and triggering further liquidations.
}

\begin{figure}[tbh]
    \centering  
    \includegraphics[width=0.95\columnwidth] {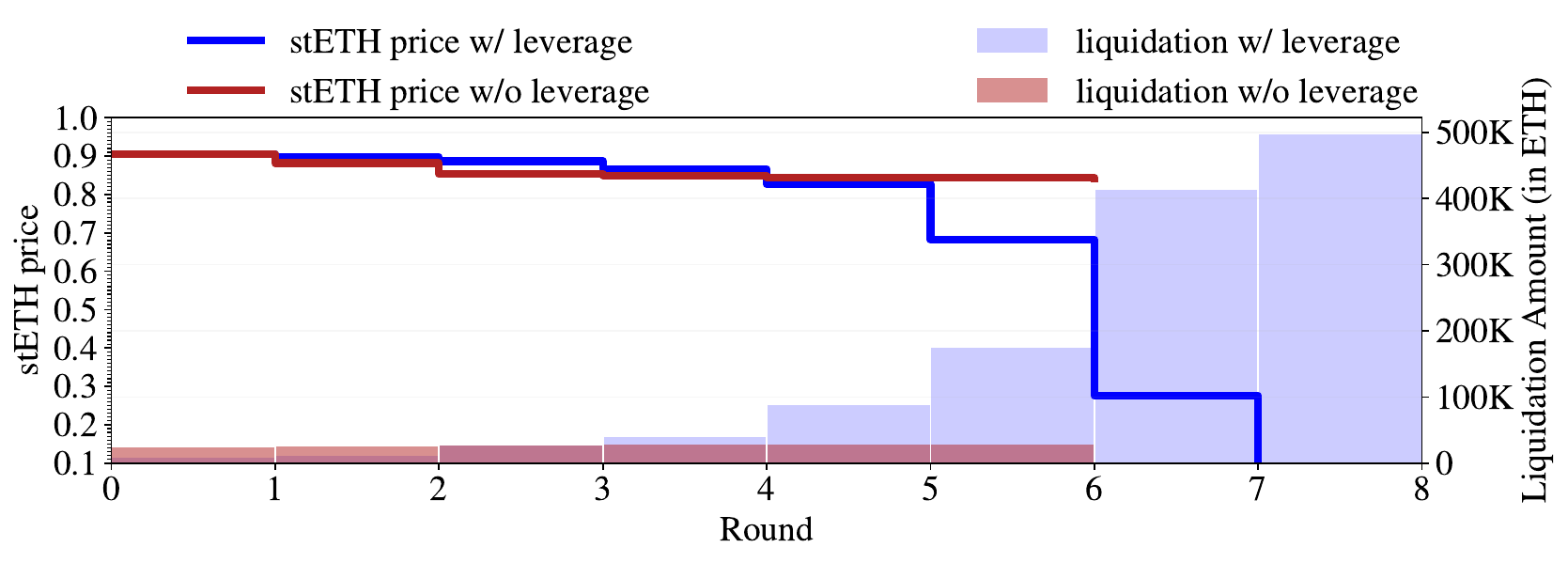}
    \caption{Comparision of \stETH price and liquidation amount with and without leverage staking. The blue and red lines (bars) show the change of the \stETH price (the change of liquidation amount) when users in \GroupLV adopt or do not adopt the leverage staking strategy.
    }
    \label{fig: cascading_liquidation_results}
\end{figure}

Figure~\ref{fig: cascading_liquidation_results} illustrates the comparative simulation results for scenarios where users adopt or do not adopt the leverage staking strategy. In the absence of leverage staking, the \stETH price stabilized at $0.84$~\ETH in the last rounds, leading to a comparatively modest liquidation amount of $28{,}201$~\ETH. However, with the application of leverage staking, the \stETH price plummeted to $0.01$~\ETH at the end of the simulation, and the liquidation amount ($497{,}375$~\ETH) escalated to 16 times that of the scenario where such strategies were not applied. Our simulation findings indicate that adopting leverage staking strategies significantly exacerbates the risk of cascading liquidation in response to market downturns. This underscores the importance of robust risk management within the \LSD system. Implementing such strategies without considering their potential impact on market stability can lead to adverse outcomes that affect a wide range of stakeholders.

\subsubsection{SQ3}
SQ3 aims to explore the impact of leverage staking positions on ordinary positions during significant \stETH devaluation. This simulation constructs two scenarios: a control scenario, which involves only the ordinary group (\GroupOD) consisting of 442 users, and an experimental scenario, where \GroupLV (442 users) and \GroupOD (442 users) coexist on the Aave platform. Both scenarios are subjected to identical simulation processes to record the number of liquidated positions within \GroupOD and the fluctuations in the \stETH price, allowing for a direct comparison of outcomes with and without the influence of leverage staking.

\smallskip
\insightbox { \label{insights-3}
\normalfont 
Leverage staking introduces broader systemic risks because it significantly exacerbates the liquidation of ordinary positions during \stETH devaluation.
}

\begin{figure}[h]
\centering

\begin{minipage}{.495\textwidth}
  \centering
   \includegraphics[width =\columnwidth]
{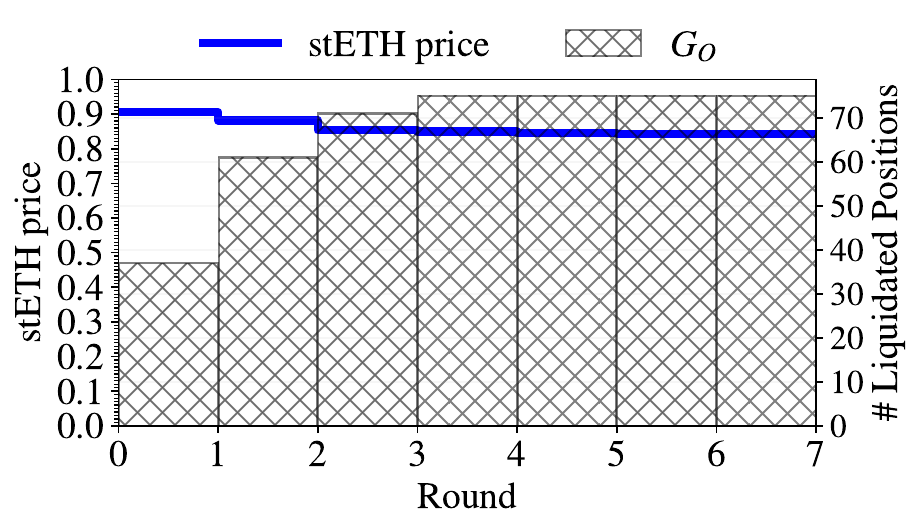}
  \caption{The change of the \stETH price and \#~liquidated positions in \GroupOD in the absence of the leverage staking group (\GroupLV).}
\label{all_case2_liquidatable_ordinary_users_without_leverage}
\end{minipage}%
\hfill
\begin{minipage}{.495\textwidth}
  \centering
  \includegraphics[width =\columnwidth]
{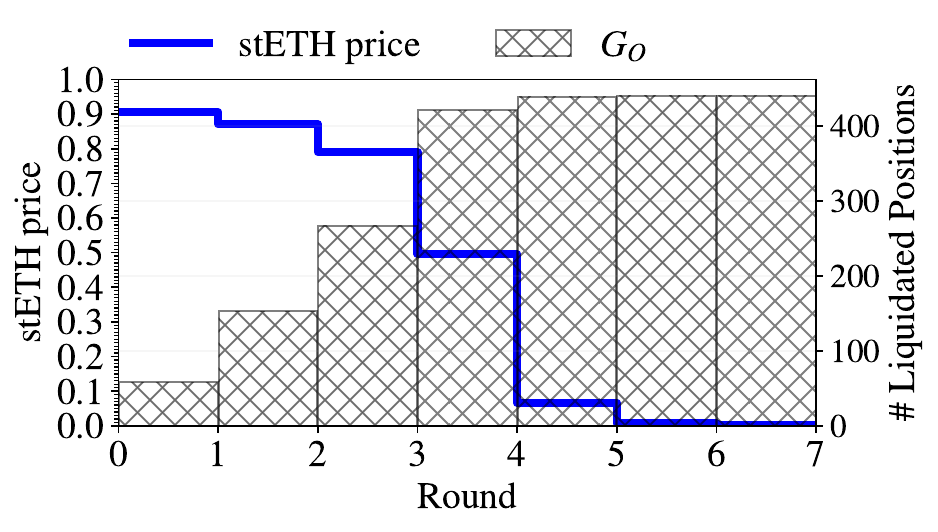}
\caption{The change of \stETH price and \#~liquidated positions in \GroupOD in the presence of the leverage staking group (\GroupLV).}
\label{all_case2_liquidatable_ordinary_users_with_leverage}
\end{minipage}
\end{figure}

The simulation results for the control scenario (see Figure~\ref{all_case2_liquidatable_ordinary_users_without_leverage}) indicate that $75$ ordinary positions are liquidated in the absence of \GroupLV. In contrast, the results for the experimental scenario (see Figure~\ref{all_case2_liquidatable_ordinary_users_with_leverage}) reveal that $440$ ordinary positions are liquidated, suggesting a significant increase in liquidations when \GroupLV is present. Our simulation results suggest that leverage staking not only intensifies the risk profile of individual portfolios but also contributes to broader systemic risks, particularly during periods of sharp declines in \LSD prices. The comparative analysis of liquidation rates between the two scenarios underscores the influence that leverage staking positions can exert on the stability of ordinary positions. 

The increased liquidations in the presence of \GroupLV imply a contagion effect, where vulnerabilities in leveraged positions can cascade to affect even traditionally less risky, ordinary positions.  This suggests that systems designed to stabilize market dynamics need to account not only for individual positions but also for their interdependencies. Therefore, our simulation results highlight the necessity for regulatory frameworks and platform governance structures to consider these interconnections.

\subsubsection{SQ4}
As discussed in Section~\ref{sec:liquidation-risk}, users holding leverage staking positions might choose to deleverage during a decline in \stETH value. SQ4 is designed to examine the effects of such deleveraging actions on the \stETH price and \LSD market participants. We simulate two scenarios: the control scenarios where \GroupLV does not deleverage and the experimental scenarios where \GroupLV chooses to deleverage at the beginning of the simulation (round 0).

\begin{figure}[thb]
\begin{minipage}{.495\textwidth}
  \centering
   \includegraphics[width =\columnwidth]
{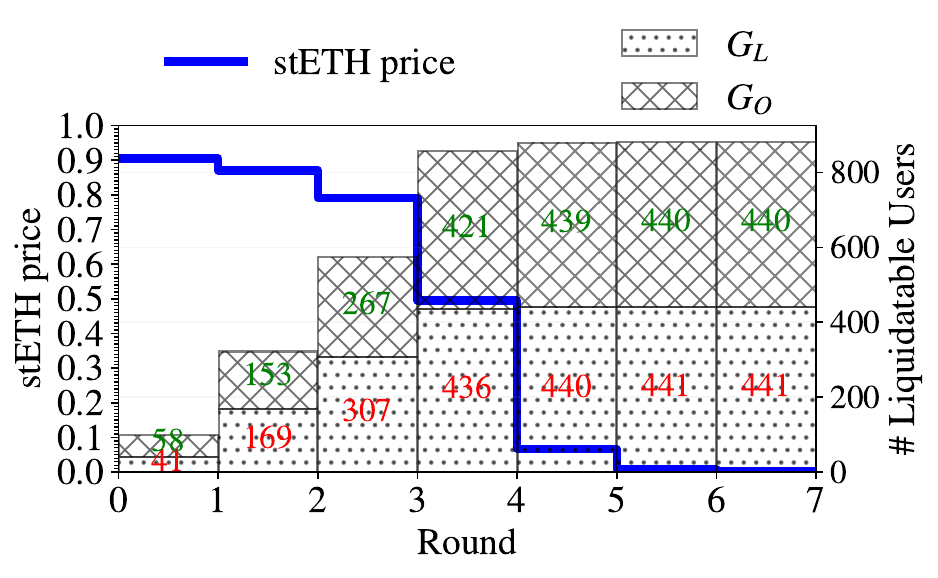}
  \caption{The change of \stETH price and \#~liquidated positions without deleveraging.} 
\label{all_cascading_liquidation_results_case4_liquidatable_users}
\end{minipage}%
\hfill
\begin{minipage}{.495\textwidth}
  \centering
  \includegraphics[width =\columnwidth]
{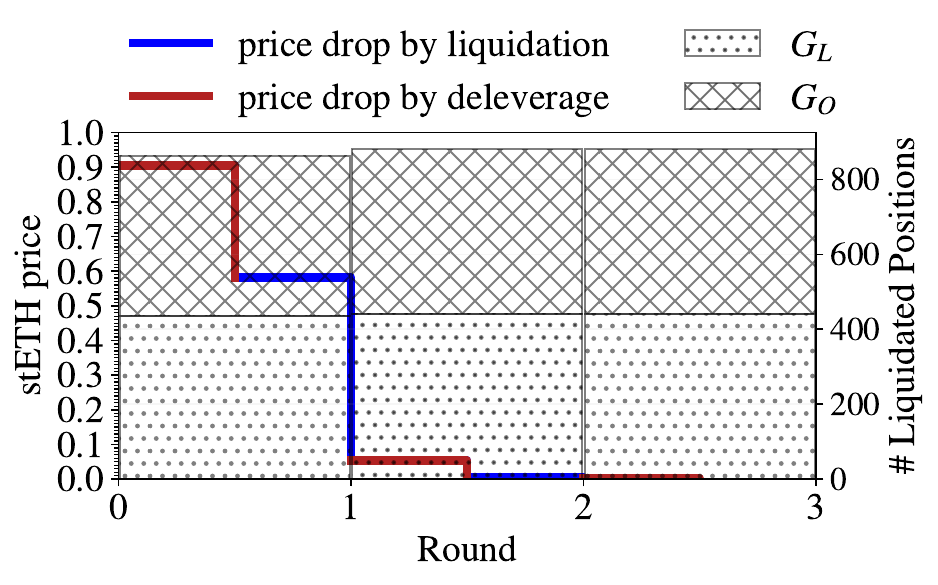}
\caption{The change of \stETH price and \#~liquidated positions with deleveraging.}
\label{all_case3_liquidatable_users_with_leverage_and_deleverage}
\end{minipage}
\end{figure}

\smallskip
\insightbox { \label{insights-4}
\normalfont 
The deleveraging actions taken by \GroupLV can introduce additional selling pressure, thereby intensifying the liquidation cascades among system participants.
}

When users in \GroupLV do not choose to deleverage, our simulation result (see~Figure~\ref{all_cascading_liquidation_results_case4_liquidatable_users}) shows that the liquidation ends in $7$ round, with $441$ users in \GroupLV and $440$ users in \GroupOD being liquidated. Conversely, when \GroupLV decides to deleverage at round 0, the liquidation process is significantly shortened, ending in just 3 rounds.  This pronounced difference underscores the critical role that deleveraging actions play in market dynamics. Not only do they shorten the duration of liquidations, but they also potentially amplify market volatility by introducing additional selling pressure. This indicates that deleveraging action can exacerbate systemic risk by accelerating the liquidation cascade, affecting a broader range of system participants.

\section{Discussion and Future Research Directions} 

Our comprehensive stress tests on the Lido-Aave-Curve LSD ecosystem reveal critical vulnerabilities and dynamic interplays under extreme conditions of significant \stETH devaluation. These simulations illustrate that leverage staking strategies, while innovative, expose the market to heightened risks.  The presence of leverage staking significantly escalates the risk of cascading liquidations within the \LSD ecosystem. This finding underlines a crucial concern: systemic risk is exacerbated not only through direct liquidations but also via the market pressures these actions generate. The selling pressure on \stETH, driven by both liquidations and deleveraging actions, can trigger a contagion effect across the system, further depressing \stETH prices and adversely impacting the financial stability of broader market participants, including ordinary users. Therefore, it is crucial to strike a balance between leveraging opportunities for higher returns and the potential for destabilization in LSD ecosystems.

\smallskip
Building upon these insights, future research can pursue several avenues. A crucial direction is the development of refined models that simulate a broader range of conditions, incorporating more granular behaviors of market participants and liquidity variations. This could lead to more robust parameterization of platforms such as Aave, similar to the ``safe parameterization'' design used in traditional finance, which aims to mitigate risks without stifling innovation. Additionally, exploring new regulatory frameworks tailored to \LSDs could help mitigate the systemic shocks observed in our simulations. By integrating advanced risk management strategies and regulatory innovations, future research can contribute to creating a more resilient \LSD ecosystem. This involves a holistic approach to understanding the interdependencies and collective behaviors that drive the resilience of these platforms.

\section{Conclusion}
This work systematically examines the leverage staking strategy with \LSDs, providing both analytical and empirical insights into its opportunities and risks. Our analytical study introduces a formal model that captures both direct and indirect leverage staking strategies, establishing key metrics such as leverage amounts, loops, multipliers, and APRs. Our empirical study builds heuristics to identify historical leverage staking positions and evaluate their performance. Notably, we observe that the majority of these positions yield an \APR exceeding that of conventional staking, highlighting their high-return potential.

However, these financial opportunities come with inherent risks. To address this, we conduct stress tests under extreme market scenarios to assess the vulnerabilities of leverage staking. Our findings reveal that leverage staking significantly amplifies the risk of cascading liquidations, leading to intensified selling pressure and systemic instability. Additionally, it propagates a contagion effect, exacerbating the liquidation of ordinary positions and posing broader systemic risks to the \LSD ecosystem.

We hope this work serves as a foundation for academic researchers and protocol developers to design robust risk assessment frameworks and implement safe parameterizations, ultimately enhancing the resilience and sustainability of the \LSD ecosystem.


\clearpage
\bibliographystyle{ieeetr}
\bibliography{references}

\begin{thebibliography}{10}

\bibitem{wood2014ethereum}
G.~Wood, ``Ethereum: A secure decentralised generalised transaction ledger,''
  {\em Ethereum project yellow paper}, vol.~151, pp.~1--32, 2014.

\bibitem{grandjean2023ethereum}
D.~Grandjean, L.~Heimbach, and R.~Wattenhofer, ``Ethereum proof-of-stake
  consensus layer: Participation and decentralization,'' {\em arXiv preprint
  arXiv:2306.10777}, 2023.

\bibitem{schwarz2022three}
C.~Schwarz-Schilling, J.~Neu, B.~Monnot, A.~Asgaonkar, E.~N. Tas, and D.~Tse,
  ``Three attacks on proof-of-stake ethereum,'' in {\em International
  Conference on Financial Cryptography and Data Security}, pp.~560--576,
  Springer, 2022.

\bibitem{agrawal2022proofs}
S.~Agrawal, J.~Neu, E.~N. Tas, and D.~Zindros, ``Proofs of proof-of-stake with
  sublinear complexity,'' {\em arXiv preprint arXiv:2209.08673}, 2022.

\bibitem{tang2023transaction}
W.~Tang and D.~D. Yao, ``Transaction fee mechanism for proof-of-stake
  protocol,'' {\em arXiv preprint arXiv:2308.13881}, 2023.

\bibitem{buterin2020combining}
V.~Buterin, D.~Hernandez, T.~Kamphefner, K.~Pham, Z.~Qiao, D.~Ryan, J.~Sin,
  Y.~Wang, and Y.~X. Zhang, ``Combining ghost and casper,'' {\em arXiv preprint
  arXiv:2003.03052}, 2020.

\bibitem{neu2021ebb}
J.~Neu, E.~N. Tas, and D.~Tse, ``Ebb-and-flow protocols: A resolution of the
  availability-finality dilemma,'' in {\em 2021 IEEE Symposium on Security and
  Privacy (SP)}, pp.~446--465, IEEE, 2021.

\bibitem{cong2022staking}
L.~W. Cong, Z.~He, and K.~Tang, ``Staking, token pricing, and crypto carry,''
  {\em Available at SSRN 4059460}, 2022.

\bibitem{chitra2021competitive}
T.~Chitra, ``Competitive equilibria between staking and on-chain lending,''
  2021.

\bibitem{tzinas2023principal}
A.~Tzinas and D.~Zindros, ``The principal--agent problem in liquid staking,''
  {\em Cryptology ePrint Archive}, 2023.

\bibitem{scharnowski2023economics}
S.~Scharnowski and H.~Jahanshahloo, ``The economics of liquid staking
  derivatives: Basis determinants and price discovery,'' {\em Available at SSRN
  4180341}, 2023.

\bibitem{cintra2023detecting}
T.~N. Cintra and M.~P. Holloway, ``Detecting depegs: Towards safer passive
  liquidity provision on curve finance,'' {\em arXiv preprint
  arXiv:2306.10612}, 2023.

\bibitem{heimbach2023defi}
L.~Heimbach, E.~Schertenleib, and R.~Wattenhofer, ``Defi lending during the
  merge,'' in {\em 5th Conference on Advances in Financial Technologies}, 2023.

\bibitem{wang2022speculative}
Z.~Wang, K.~Qin, D.~V. Minh, and A.~Gervais, ``Speculative multipliers on defi:
  Quantifying on-chain leverage risks,'' {\em Financial Cryptography and Data
  Security}, 2022.

\bibitem{bitcoin}
S.~Nakamoto, ``Bitcoin: A peer-to-peer electronic cash system,'' 2008.
\newblock Available at: \url{https://bitcoin.org/bitcoin.pdf}.

\bibitem{buterin2013ethereum}
V.~Buterin {\em et~al.}, ``Ethereum white paper,'' {\em GitHub repository},
  vol.~1, pp.~22--23, 2013.

\bibitem{werner2022sok}
S.~Werner, D.~Perez, L.~Gudgeon, A.~Klages-Mundt, D.~Harz, and W.~Knottenbelt,
  ``Sok: Decentralized finance (defi),'' in {\em Proceedings of the 4th ACM
  Conference on Advances in Financial Technologies}, pp.~30--46, 2022.

\bibitem{daian2019snow}
P.~Daian, R.~Pass, and E.~Shi, ``Snow white: Robustly reconfigurable consensus
  and applications to provably secure proof of stake,'' in {\em International
  Conference on Financial Cryptography and Data Security}, pp.~23--41,
  Springer, 2019.

\bibitem{gavzi2019proof}
P.~Ga{\v{z}}i, A.~Kiayias, and D.~Zindros, ``Proof-of-stake sidechains,'' in
  {\em 2019 IEEE Symposium on Security and Privacy (SP)}, pp.~139--156, IEEE,
  2019.

\bibitem{kiayias2017ouroboros}
A.~Kiayias, A.~Russell, B.~David, and R.~Oliynykov, ``Ouroboros: A provably
  secure proof-of-stake blockchain protocol,'' in {\em Annual International
  Cryptology Conference}, pp.~357--388, Springer, 2017.

\bibitem{bano2019sok}
S.~Bano, A.~Sonnino, M.~Al-Bassam, S.~Azouvi, P.~McCorry, S.~Meiklejohn, and
  G.~Danezis, ``Sok: Consensus in the age of blockchains,'' in {\em Proceedings
  of the 1st ACM Conference on Advances in Financial Technologies},
  pp.~183--198, 2019.

\bibitem{EthMerge2022}
``{The Merge},'' 2023.
\newblock Available at: \url{https://ethereum.org/en/roadmap/merge/}.

\bibitem{daian2020flash}
P.~Daian, S.~Goldfeder, T.~Kell, Y.~Li, X.~Zhao, I.~Bentov, L.~Breidenbach, and
  A.~Juels, ``Flash boys 2.0: Frontrunning in decentralized exchanges, miner
  extractable value, and consensus instability,'' in {\em 2020 IEEE Symposium
  on Security and Privacy (SP)}, pp.~910--927, IEEE, 2020.

\bibitem{liu2022empirical}
Y.~Liu, Y.~Lu, K.~Nayak, F.~Zhang, L.~Zhang, and Y.~Zhao, ``Empirical analysis
  of eip-1559: Transaction fees, waiting times, and consensus security,'' in
  {\em Proceedings of the 2022 ACM SIGSAC Conference on Computer and
  Communications Security}, pp.~2099--2113, 2022.

\bibitem{EthHistory2023}
Ethereum.org, ``The history of ethereum,'' 2023.
\newblock Available at: \url{https://ethereum.org/en/history/}.

\bibitem{lido-tokens-integration-guide2023}
``Lido tokens integration guide,'' 2023.
\newblock Available at:
  \url{https://docs.lido.fi/guides/lido-tokens-integration-guide}.

\bibitem{liu2023anatomy}
J.~Liu, I.~Makarov, and A.~Schoar, ``Anatomy of a run: The terra luna crash,''
  tech. rep., National Bureau of Economic Research, 2023.

\bibitem{lee2023dissecting}
S.~Lee, J.~Lee, and Y.~Lee, ``Dissecting the terra-luna crash: Evidence from
  the spillover effect and information flow,'' {\em Finance Research Letters},
  vol.~53, p.~103590, 2023.

\end{thebibliography}

\clearpage
\appendix
\section{Aave Parameter Configuration}
Table~\ref{tab:aave_params} depicts the historical changes of Aave parameter configurations. We crawl the \href{https://github.com/aave/protocol-v2/blob/ce53c4a8c8620125063168620eba0a8a92854eb8/contracts/interfaces/ILendingPoolConfigurator.sol\#L80}{\texttt{collateralConfigurationChanged}} events for Aave V$2$ lending pool.

\begin{table}[tbh]
    \centering
    \resizebox{\textwidth}{!}{
    \begin{tabular}{clll}
    \toprule
     Block Number & \multicolumn{1}{c}{Transaction Hash} & LTV($l$) & LT \\
     \midrule
     14289297 & \href{https://etherscan.io/tx/0x94780dc1914af5aec3b6d303e2d974669074bbceb6d1baac7d93ad0400593db0}{0x94780dc1914af5aec3b6d303e2d974669074bbceb6d1baac7d93ad0400593db0} & 0.70 & 0.75\\
      14693506 & \href{https://etherscan.io/tx/0x993926559f17a579c3b0c0b5fc83fd3c9d5772e9b314f7f0e78e704c6b984726}{0x993926559f17a579c3b0c0b5fc83fd3c9d5772e9b314f7f0e78e704c6b984726} & 0.73 &0.75\\
 14804760 & \href{https://etherscan.io/tx/0xbc40546b65ada9f5d4f8346f405a5f9c0da6d8f66bb27b7c64c0efa70eeae080}{0xbc40546b65ada9f5d4f8346f405a5f9c0da6d8f66bb27b7c64c0efa70eeae080} & 0.69 &0.81\\
 14837221 & \href{https://etherscan.io/tx/0x5206c144845ac63f982125f51c31b0fc474655281e2d433f8270505d87f8bbf4}{0x5206c144845ac63f982125f51c31b0fc474655281e2d433f8270505d87f8bbf4} & 0.70 &0.75\\
 14999895& \href{https://etherscan.io/tx/0x25e33e04d2e5d92acd91f542a2677045e59b2d6b385e4fa0315a404396bc5c99}{0x25e33e04d2e5d92acd91f542a2677045e59b2d6b385e4fa0315a404396bc5c99} & 0.69 & 0.81\\
 15759644 & \href{https://etherscan.io/tx/0x48653014a79433bf5f21781424b306a697f4476077a8b92e17b2ae6eda58706e}{0x48653014a79433bf5f21781424b306a697f4476077a8b92e17b2ae6eda58706e} & 0.72 & 0.83\\
 16392718 & \href{https://etherscan.io/tx/0xdcaa33ddef0700a2a625e0b7a0c2da14499901e42c073064390f11dd83cefd19}{0xdcaa33ddef0700a2a625e0b7a0c2da14499901e42c073064390f11dd83cefd19} & 0.69 & 0.81\\
     \bottomrule
    \end{tabular}
    }
    \caption{Historical changes of Aave V$2$ parameter configurations.}
    \label{tab:aave_params}
\end{table}

\section{Generalized Formalization For Leverage Staking}\label{appx:generalized_formalization}

\subsection{Generalized Formalization}
In addition to the standardized cases discussed in Section~\ref{sec:analytical}, real-world leverage lending situations can exhibit substantial variation among users. Specifically, we delineate the following variations using the direct leverage staking strategy as an example.

\emph{(i)} Within each leverage staking loop, \user may choose not to supply all the \stETH acquired from Lido as collateral on Aave. Instead, in the $k^{\text{th}}$ loop, \user may opt to supply only $c_k$ ($c_k\in[0,1]$) percent of the \stETH amount. \emph{(ii)} In the $k^{\text{th}}$ loop, \user has the option to borrow an amount of \ETH that is less than the maximum borrowing capacity. In this scenario, \user's effective borrowing capacity becomes $b_k\cdot l$, where $b_k\in[0,1]$. \emph{(iii)} In the $k^{\text{th}}$ loop, \user has the flexibility to restake part of the borrowed \ETH on Lido. \user may choose to restake only $s_k$ ($s_k\in[0,1]$) percent of the \ETH borrowed in the the $(k-1)^{\text{th}}$ loop (or the principal amount when $k=1$) at the start of the $k^{\text{th}}$ loop. After borrowing \stETH from Aave, \user may restake only $s_{k+1}$ ($s_{k+1}\in[0,1]$) percent of the borrowed \ETH at the end of the $k^{\text{th}}$ loop (equals to the beginning of $(k+1)^{\text{th}}$ loop). Note that as illustrated by Figure~\ref{fig: lev_loop}, the stake and restake parameters ($s_k$ and $s_{k+1}$) together establish the the $k^{\text{th}}$ loop. Equation~\ref{eq:generalized} introduces a generalized formalization to accommodate these variations. 
\begin{equation} \label{eq:generalized}
  {\footnotesize
    \begin{split}
    & A_{(S,n)} = S\cdot\sum_{k=1}^{n+1}  \left(\prod_{i=1}^k s_i\right) \cdot \left(\prod_{i=1}^{i-1} c_i\right)  \cdot \left(\prod_{i=1}^{k-1} b_i\right) \cdot \left(\frac{l\cdot p_{t_0}^{a}}{p_{t_0}^{m}}\right)^{k-1}\\
    & C_{(S,n)} = S\cdot\sum_{k=1}^n \left(\prod_{i=1}^k s_i\right) \cdot \left(\prod_{i=1}^k c_i\right)  \cdot \left(\prod_{i=1}^{k-1} b_i\right) \cdot \frac{(l\cdot p_{t_0}^{a})^{k-1}}{(p_{t_0}^{m})^{k}} \\
    & B_{(S,n)} = S\cdot\sum_{k=1}^{n} \left(\prod_{i=1}^k s_i\right) \cdot \left(\prod_{i=1}^{k} c_i\right)\cdot \left(\prod_{i=1}^{k} b_i\right) \cdot\left(\frac{l\cdot p_{t_0}^{a}}{p_{t_0}^{m}}\right)^{k}  \\
    &  HF_{\mathsf{U_i}}(p_{t_c}^{a}|p_{t_0}^{a}) =\frac{C_{(S,n)}\cdot p_{t_c}^{a}\cdot \text{LT}}{B_{(S,n)}} \\
    \end{split}
    }
\end{equation}

\section{Leverage Staking Detection Algorithm} \label{appx:algo}

Algorithms~\ref{alg: leverage-staking-detection} and \ref{alg: indirect-leverage-staking-detection} depict the heuristics used to detect addresses that have performed direct and indirect leverage staking respectively.

\RestyleAlgo{ruled}
\SetKwComment{Comment}{// }{}
\SetNoFillComment
\SetKw{Continue}{continue}
\SetKwInOut{Parameter}{Param}
\begin{algorithm}[h]
{\small
\caption{Direct Leverage Staking Detection.}\label{alg: leverage-staking-detection}
\KwIn{ An address $\mathsf{addr}$
}
\KwOut{
$\mathsf{addr}$'s leverage staking actions.
}

Extract $\mathsf{addr}$'s \texttt{deposit} events $\{w_i\}_i$ and \texttt{borrow} events $\{b_j\}_j$ on Aave, and \texttt{stake} events $\{s_k\}_k$ on Lido;

Let $\mathcal{E} = \{w_i\}_i \bigcup \{b_j\}_j \bigcup \{s_k\}_k$;

Convert $\mathcal{E}$ to a sequence $\mathcal{E}_s$ by sorting $\mathcal{E}$ in chronological order;

\uIf{$\mathcal{E}_s$ contains a sub-sequence with a order of $(\texttt{stake}$, $\texttt{deposit}$, $\texttt{borrow}$, $\texttt{stake})$}{
\uIf{For the sub-sequence $(\texttt{stake}_0$, $\texttt{deposit}$, $\texttt{borrow}$, $\texttt{stake}_1)$: \emph{(i)} The \stETH amount received in $\texttt{stake}_0$ event $\approx$  the \stETH amount in $\texttt{deposit}$ event; \emph{(ii)} The \stETH amount in $\texttt{deposit}$ event $>$ the \ETH amount in $\texttt{borrow}$ event;  \emph{(iii)} The \ETH amount in $\texttt{borrow}$ event $\approx$ the \ETH amount in $\texttt{stake}_1$ event
}{
\Return $\mathcal{E}_s$;
}
\Return $\emptyset$;
}
\Else{\Return $\emptyset$;} 
}
\end{algorithm}

\RestyleAlgo{ruled}
\SetKwComment{Comment}{// }{}
\SetNoFillComment
\SetKw{Continue}{continue}
\SetKwInOut{Parameter}{Param}
\begin{algorithm}[h]
{\small
\caption{Indirect Leverage Staking Detection.}\label{alg: indirect-leverage-staking-detection}
\KwIn{ An address $\mathsf{addr}$
}
\KwOut{
$\mathsf{addr}$'s indirect leverage staking actions.
}

Extract $\mathsf{addr}$'s \texttt{deposit} events $\{w_i\}_i$ and \texttt{borrow} events $\{b_j\}_j$ on Aave, and \texttt{swap} events $\{s_k\}_k$ on Curve;

Let $\mathcal{E} = \{w_i\}_i \bigcup \{b_j\}_j \bigcup \{s_k\}_k$;

Convert $\mathcal{E}$ to a sequence $\mathcal{E}_s$ by sorting $\mathcal{E}$ in chronological order;

\uIf{$\mathcal{E}_s$ contains a sub-sequence with a order of $(\texttt{swap}$, $\texttt{deposit}$, $\texttt{borrow}$, $\texttt{swap})$}{
\uIf{For the sub-sequence $(\texttt{swap}_0$, $\texttt{deposit}$, $\texttt{borrow}$, $\texttt{swap}_1)$: \emph{(i)} The \stETH amount received in $\texttt{swap}_0$ event $\approx$  the \stETH amount in $\texttt{deposit}$ event; \emph{(ii)} The \stETH amount in $\texttt{deposit}$ event $>$ the \ETH amount in $\texttt{borrow}$ event;  \emph{(iii)} The \ETH amount in $\texttt{borrow}$ event $\approx$ the \ETH amount in $\texttt{swap}_1$ event
}{
\Return $\mathcal{E}_s$;
}
\Return $\emptyset$;
}
\Else{\Return $\emptyset$;} 
}
\end{algorithm}

\end{document}